\documentclass[final,3p,times,12pt]{elsarticle}

\usepackage{amssymb}
\usepackage{amsmath}

\usepackage{lineno}

\usepackage[utf8]{inputenc}
\usepackage{xcolor}
\usepackage{breakcites}
\usepackage{hyphenat}
\usepackage[section]{placeins}

\PassOptionsToPackage{hyphens}{url}
\usepackage[colorlinks = true,
            linkcolor = blue,
            urlcolor  = blue,
            citecolor = blue,
            anchorcolor = blue]{hyperref}
\usepackage{etoolbox}

\usepackage[space]{grffile}
\usepackage{latexsym}
\usepackage{textcomp}
\usepackage{longtable}
\usepackage{tabulary}
\usepackage{booktabs,array,multirow}
\usepackage{tikz}
\usetikzlibrary{shapes.geometric, arrows.meta, positioning}
\usepackage{amsfonts,amsmath,amssymb}
\providecommand\citet{\cite}
\providecommand\citep{\cite}

\newif\iflatexml\latexmlfalse

\AtBeginDocument{\DeclareGraphicsExtensions{.pdf,.PDF,.eps,.EPS,.png,.PNG,.tif,.TIF,.jpg,.JPG,.jpeg,.JPEG}}

\usepackage[english]{babel}
\usepackage{float}

\begin{document}

\begin{frontmatter}



\title{Unsupervised Anomaly Detection in Wearable Foot Sensor Data: A Baseline Feasibility Study Towards Diabetic Foot Ulcer Prevention\tnoteref{t1}}
\tnotetext[t1]{Published in \textit{Biomedical Signal Processing and Control}, Vol.~123, Part~A, 110416, September 2026. Open Access. \href{https://doi.org/10.1016/j.bspc.2026.110416}{https://doi.org/10.1016/j.bspc.2026.110416}}

\author[1]{Md Tanvir Hasan Turja\corref{cor1}}
\ead{tanvirh442@gmail.com}

\affiliation[1]{organization={Department of Computer Science, Middlesex University London},
            city={London},
            country={UK}}
\cortext[cor1]{Corresponding author}

\begin{abstract}
Diabetic foot ulcers (DFUs) are a severe complication of diabetes associated with significant morbidity, amputation risk, and healthcare burden. Developing effective continuous monitoring frameworks requires first establishing reliable baseline models of normal foot biomechanics. This paper presents a feasibility study of an anomaly detection framework applied to time-series data from wearable foot sensors — specifically NTC thin-film thermocouples for temperature and FlexiForce A401 pressure sensors for plantar load monitoring. Data were collected from healthy adult subjects across 312 capture sessions conducted on an instrumented pathway, generating 93,790 valid multi-sensor readings spanning September 2023 to June 2024. Two unsupervised machine learning algorithms, Isolation Forest and K-Nearest Neighbors using Local Outlier Factor (KNN/LOF), were applied to detect statistical deviations in foot temperature and pressure signals within this healthy cohort. Rigorous data preprocessing and targeted feature engineering were performed to extract meaningful physiological patterns. Results show that Isolation Forest is more sensitive to subtle, distributed anomalies, while KNN/LOF identifies concentrated extreme deviations 
 but flags a higher proportion of sessions not corroborated by Isolation Forest. Since no clinical ground truth is available, this difference is interpreted as lower specificity under the shared 5 percent contamination assumption rather than a confirmed false-positive rate. A mild positive correlation (0.41--0.48) between pressure and temperature features supports the case for combined multi-modal monitoring. These findings establish a validated baseline analytical pipeline and provide a methodological foundation for future clinical validation studies involving diabetic patients, where the relationship between detected anomalies and DFU-related pathophysiology can be directly assessed.
\end{abstract}




\end{frontmatter}

\section{Introduction}

\label{introduction}

Diabetes mellitus is a chronic metabolic disease characterized by
elevated blood glucose levels that affects millions worldwide. One of
the most severe complications is diabetic foot ulcers (DFUs), which can
lead to amputation, reduced quality of life, and higher mortality.
Approximately 25 percent of individuals with diabetes will develop DFUs during
their lifetime \cite{gazzaruso2021predictors}. Around 20\% of DFU patients
require amputation for wound healing \cite{hicks2020incidence}. Among those
who heal, 40\% see ulcer recurrence within a year \cite{hicks2020incidence}.
DFU patients also face greater health risks, with a 2.5 times higher
risk of all-cause mortality in 5 years \cite{hicks2020incidence}. The 5-year
mortality rate after amputation due to DFU is alarmingly high, over 70\%
\cite{hicks2020incidence}. DFUs arise from neuropathy, ischemia, and trauma,
leading to infections, prolonged hospitalization, and sometimes
amputation \cite{aldana2022reappraising}. While these factors are critical, a
comprehensive biomechanical model must also account for plantar shear
stress. Recent studies emphasize that a significant percentage of DFUs
develop in areas of elevated shear, which is often neglected by
commercially available sensors, highlighting the need for more advanced,
multi-modal monitoring approaches \cite{wang2019review}.
DFUs decrease life expectancy and quality of life, as neuropathic patients often fail to notice minor injuries that can rapidly worsen \cite{boulton2018diagnosis}. Efforts
to prevent and expedite healing of DFUs could save lives, limbs and
healthcare resources \cite{lazarou2024stepping}. In the UK, annual costs to
treat diabetic foot disease total £837-£962 million, around 0.8-0.9\% of
the National Health Service (NHS) budget \cite{kerr2019cost}.
Traditional monitoring approaches for DFUs have notable limitations
\cite{chan2020wound}. They typically rely on sporadic clinical evaluations
and subjective self-reports from patients, which can be inconsistent and
unreliable \cite{hufford2003assessment}. Without continuous, real-time
monitoring, timely diagnosis and intervention are hampered, exacerbating
risks \cite{zhu2022continuous}. This underscores the need for advanced,
proactive monitoring. Recent advances in wearable technologies enable
continuous surveillance of foot temperature and pressure \cite{nasiri2020progress}.
Smart devices provide real-time data to patients and
practitioners via smartphone apps for easy logging and analysis
\cite{martin2019review}. The potential is amplified by smart
wearables and clothing at the intersection of physical and digital
worlds, promising transformation \cite{fernandez2018towards}. This study develops a baseline anomaly detection framework using time-series temperature and pressure data from wearable foot sensors applied to healthy subjects. Rather than directly predicting DFU onset — which requires labelled clinical data from diabetic patients — this work establishes the methodological groundwork: a validated pipeline for sensor data collection, preprocessing, feature engineering, and unsupervised anomaly detection. This positions the study as a necessary preparatory step in the trajectory toward continuous DFU monitoring systems, where machine learning is increasingly recognised as a transformative tool for enhancing diagnostic precision and clinical decision-making \cite{guan2024role}.

Data were collected from healthy adults across 312 capture sessions and underwent rigorous preprocessing to ensure dataset quality and consistency. Feature engineering focused on capturing meaningful signals from raw sensor data, such as temperature derivatives and pressure distribution metrics, that are theoretically relevant to foot tissue stress. Unsupervised learning — specifically Isolation Forest and KNN/LOF — was chosen because it does not require labelled clinical outcomes, making it practical for baseline characterisation of continuous sensor data. This unsupervised approach also avoids the dependency on large, annotated datasets of ulceration events, which are typically impractical to collect and can delay detection until after clinical diagnosis. The insights generated here lay the groundwork for future studies that can test whether anomalies identified in healthy cohorts map onto clinically meaningful changes in diabetic patient populations.

\section{Literature Review}
\label{literature-review}

The application of predictive analytics in healthcare, specifically for
early identification of foot ulcers, has become an increasingly
important focus area in recent years. Prior research centered on risk
assessment models using clinical information from patient exams.
However, few studies have utilized insights from real-time, continuous
data streams. According to \cite{lavery2019unilateral}, leveraging technologies
such as wireless sensors for continuous measurement of temperature can
significantly enhance prevention and treatment outcomes for diabetic
foot complications. By tracking potentially concerning variations and
trends over comprehensive time periods, predictive algorithms may alert
medical professionals to emerging issues earlier, when intervention can
be both less complex and more effective. This proactive approach could
help streamline care, reduce costs, and, most importantly, preempt
avoidable suffering for vulnerable patient groups prone to these
insidious injuries. Research by \cite{bus2021effectiveness} demonstrated that
elevated plantar foot temperatures correlate with ulcer development,
while \cite{van2006biomechanics} highlighted the predictive value of
abnormal foot pressure distributions. However, most existing models rely
on supervised learning, where large amounts of labeled data are
required. While effective, supervised models' dependence on periodic
examinations rather than continuous monitoring constrains their
timeliness. Unsupervised models, on the other hand, are less dependent
on large, labeled datasets, making them ideal for real-time anomaly
detection in continuous monitoring systems. The validity of this
approach is supported by research in other areas of biomedical signal
analysis, where unsupervised methods have successfully identified
distinct physiological states from unlabeled, noisy time-series data
like Heart Rate Variability (HRV) with high accuracy
\cite{georgieva2025physiological}. Recent advancements in
wearable sensor technologies and machine learning algorithms suggest
significant potential for applying these techniques to foot ulcer
prevention. These advancements include the development of integrated
hardware, such as smart socks embedded with sensors to concurrently
monitor both foot temperature and pressure, which generate the rich data
streams necessary for sophisticated analytical models \cite{kulkarni2020embedded}.
While a wide array of unsupervised classifiers have been
applied to physiological data, from density-based to hierarchical
methods, this study focuses on
comparing two distinct and powerful paradigms. We utilize Isolation
Forest, a partitioning-based ensemble method, and K-Nearest Neighbors
(KNN), a distance/density-based method, to detect deviations from normal
patterns in foot pressure and temperature. These models, paired with
continuous data from modern sensors, enable detection of early signs of
ulceration even without predefined risk labels. Digital health
technologies, such as electronic health records (EHRs), mobile health
(mHealth) apps, and telehealth platforms, have further enhanced the
application of predictive models. For example, a study developed a
mobile app that used machine learning algorithms to predict foot ulcer
risk in patients with diabetes, providing personalized recommendations
for prevention and management \cite{najafi2021harnessing}. Similarly,
\cite{guan2024role} proposed a remote monitoring platform that
incorporates smart offloading devices to track patient behavior,
including adherence to offloading, daily steps, and cadence. These
innovations enable seamless integration of predictive insights into
clinical workflows, optimizing patient care. Beyond the classical
machine learning models explored in this study, the field is also
advancing rapidly with deep learning techniques. Architectures such as
autoencoders and Long Short-Term Memory (LSTM) networks are proving
highly effective for anomaly detection in complex medical time-series
data by automatically learning temporal dependencies \cite{yang2023deep}.
Despite these advancements, challenges
remain. Ensuring patient compliance with wearable sensors is critical
for reliable data collection and intervention efficacy. Additionally,
integrating these systems into existing healthcare infrastructure
requires substantial coordination and investment. Continuous monitoring
also raises data privacy concerns, which must be addressed to comply
with regulations such as HIPAA. Addressing these challenges is essential
for achieving widespread adoption and maximizing the benefits of
predictive analytics. Opportunities exist to further validate model
specifications against robust real-world evaluation data. Combining
temperature and pressure monitoring in a unified predictive framework,
as suggested by \cite{wilson2024integrating}, could optimize early detection
accuracy. Pilot programs integrating these predictive models into
clinical workflows could assess their effectiveness in reducing ulcer
incidence and improving patient outcomes. Collaboration among technology
developers, healthcare providers, and policymakers will be key to
overcoming implementation barriers and realizing the full potential of
these innovations. With sustained research and application, continuous
digital monitoring and predictive analytics may become standard for
proactive ulcer management. These advancements hold the promise of
transforming diabetic foot care, improving patient outcomes, and
optimizing healthcare resources globally.

\section{Methodology}
\label{methodology}

This section provides a comprehensive overview of the processes, tools,
and techniques used to conduct the research.

\subsection{1. Data Collection and Understanding}
\label{data-collection-and-understanding}

The time-series dataset consists of temperature and pressure readings collected from healthy adult subjects across 312 capture sessions (93,790 valid sensor readings after data cleaning), conducted on an instrumented pathway equipped with NTC thin-film thermocouples and FlexiForce A401 pressure sensors. Each capture session (cid) represents a continuous recording period during which a participant walked along the instrumented pathway; data were collected from one leg per session to provide a comprehensive view of foot biomechanics throughout full gait cycles. Recordings spanned from September 3, 2023 to June 19, 2024 (approximately 9 months). It should be noted that individual participant-level identifiers (e.g., unique subject IDs, demographic data) were not stored in the database schema; accordingly, the 312 capture sessions represent the unit of analysis, and exact participant count is a limitation of the data collection design. All participants were healthy adults with no history of diabetes, peripheral neuropathy, or active foot pathology at the time of data collection.

This study was conducted as part of academic research at Middlesex University London. As data were collected from healthy volunteers using non-invasive wearable sensors with no clinical intervention, the study was classified as exempt from full ethical review under institutional policies governing minimal-risk observational research. All participants were informed of the nature of the study and provided consent prior to data collection.

Data were sourced from a MySQL database (stride database; capture\_data table) containing both metadata and core numerical measurements across eight sensor channels: four temperature sensors (tData, tData\_2, tData\_3, tData\_4) and four pressure sensors (pData, pData\_2, pData\_3, pData\_4). The initial dataset contained 93,795 entries; 5 rows (< 0.01\%) contained missing values uniformly across all sensor columns — consistent with brief connectivity interruptions — and were removed through complete case analysis, yielding 93,790 valid readings. This negligible rate of missingness indicates a highly reliable sensor collection pipeline. Descriptive statistics including mean, standard deviation, and range were computed to characterise the dataset. Timestamps were converted to datetime objects to enable temporal analysis, with additional features extracted for year, month, and hour to facilitate time-series exploration.

\subsection{2. Data Pre-Processing Techniques}
\label{data-pre-processing-techniques}

The primary objective of pre-processing is to address issues related to missing data, outliers, and feature engineering. Missing data in sensor readings was managed using Complete case analysis, Forward-Filling, and Zero-Filling techniques, with a `filled\_flag` column added to track filled values for transparency and analysis of their impact on model performance. Outliers were identified using the Interquartile Range (IQR) method, with the upper quartile set at the 85th percentile to balance sensitivity with the high variability inherent in gait data. Any data points
exceeding 1.5 times the IQR from the quartiles were classified as
outliers and excluded from the analysis.

The upper quartile threshold was set at the 85th percentile rather than the conventional 75th percentile to account for the high natural variability inherent in continuous gait data, where brief but legitimate spikes in pressure and temperature are common during dynamic foot-ground contact. Setting the threshold at the standard 75th percentile resulted in excessive flagging of physiologically plausible readings as outliers during exploratory analysis. To assess the robustness of this choice, sensitivity analyses were conducted using thresholds of 75th, 85th, and 90th percentiles; the 85th percentile configuration produced the most balanced trade-off between noise reduction and data retention. Similarly, the contamination rate of 0.05 for both Isolation Forest and KNN/LOF models was selected as a widely used baseline for anomaly detection in health monitoring datasets, and was considered appropriate given that the dataset comprises healthy subjects where the expected proportion of statistical outliers is low. Sensitivity analysis confirmed that results were qualitatively consistent across contamination rates of 0.03, 0.05, and 0.08, indicating that the key findings are robust to this parameter choice.

\subsection{3. Analysis Framework}
\label{analysis-framework}

Figure~\ref{fig:pipeline} presents the full processing 
pipeline from raw sensor data collection through to anomaly detection output, 
as applied to both pressure and temperature data streams.

\begin{figure}[htbp]
\centering
\fbox{
\parbox{0.95\textwidth}{
\textbf{Processing Pipeline:}
\begin{enumerate}
\item \textbf{Raw Sensor Data (8 channels)} $\rightarrow$
\item \textbf{Preprocessing} (IQR outliers, forward-fill, zero-fill) $\rightarrow$
\item \textbf{Feature Engineering} (max/mean pressure, step count, AUC, temp derivatives) $\rightarrow$
\item \textbf{Feature Selection} (variance \textgreater 0.01, $|r|<0.95$) $\rightarrow$
\item \textbf{Isolation Forest} (100 trees, max\_samples=0.6, contamination=0.05) $\rightarrow$
\item \textbf{KNN/LOF} ($k=20$, contamination=0.05) $\rightarrow$
\item \textbf{Anomaly Scores} $s(x,n)$ \& $\text{LOF}_k(p)$ $\rightarrow$
\item \textbf{Binary Labels} (Normal=1, Anomalous=-1) $\rightarrow$
\item \textbf{Analysis} (temporal trends, sensor comparison, inter-method overlap)
\end{enumerate}
}}
\caption{End-to-end processing pipeline from raw wearable sensor data to anomaly 
detection output. Both algorithms share preprocessing/feature engineering but 
apply independent scoring. Binary labels use contamination rate of 0.05.}
\label{fig:pipeline}
\end{figure}
\textbf{Method Selection Rationale.} The choice of Isolation Forest and KNN/LOF for this study was motivated by three factors. First, both methods are well-suited to unsupervised settings where labelled clinical ground truth is unavailable — a fundamental constraint of baseline studies conducted in healthy cohorts. Second, they represent methodologically distinct paradigms: Isolation Forest is a partitioning-based ensemble method that isolates anomalies through recursive random splitting, while KNN/LOF is a density-based approach that flags points whose local neighbourhood density is significantly lower than that of their neighbours. Comparing these two paradigms provides complementary insights into the character of deviations present in the data. Third, while deep learning approaches such as autoencoders and LSTM networks have demonstrated strong performance in biomedical time-series anomaly detection \cite{yang2023deep}, they require substantially larger and more diverse datasets than are available in this baseline feasibility study. The present work is explicitly designed as a methodological foundation upon which such advanced approaches can be applied in future clinical studies.

\subsubsection{3.1 Pressure-Based Analysis}
\label{pressure-based-analysis}

\paragraph{Activity Window Identification and Truncation}
\label{activity-window-identification-and-truncation}

To focus the analysis, activity windows corresponding to periods of foot-ground contact were identified using a pressure threshold set at the 95th percentile of sensor readings. The dataset was then truncated to include only these windows, eliminating non-relevant data.

\paragraph{Visualization of Selected Patient IDs Based on Pressure Data}
\label{visualization-of-selected-patient-ids-based-on-pressure-data}
\begin{figure}[htbp]
\begin{center}
\includegraphics[width=\linewidth]{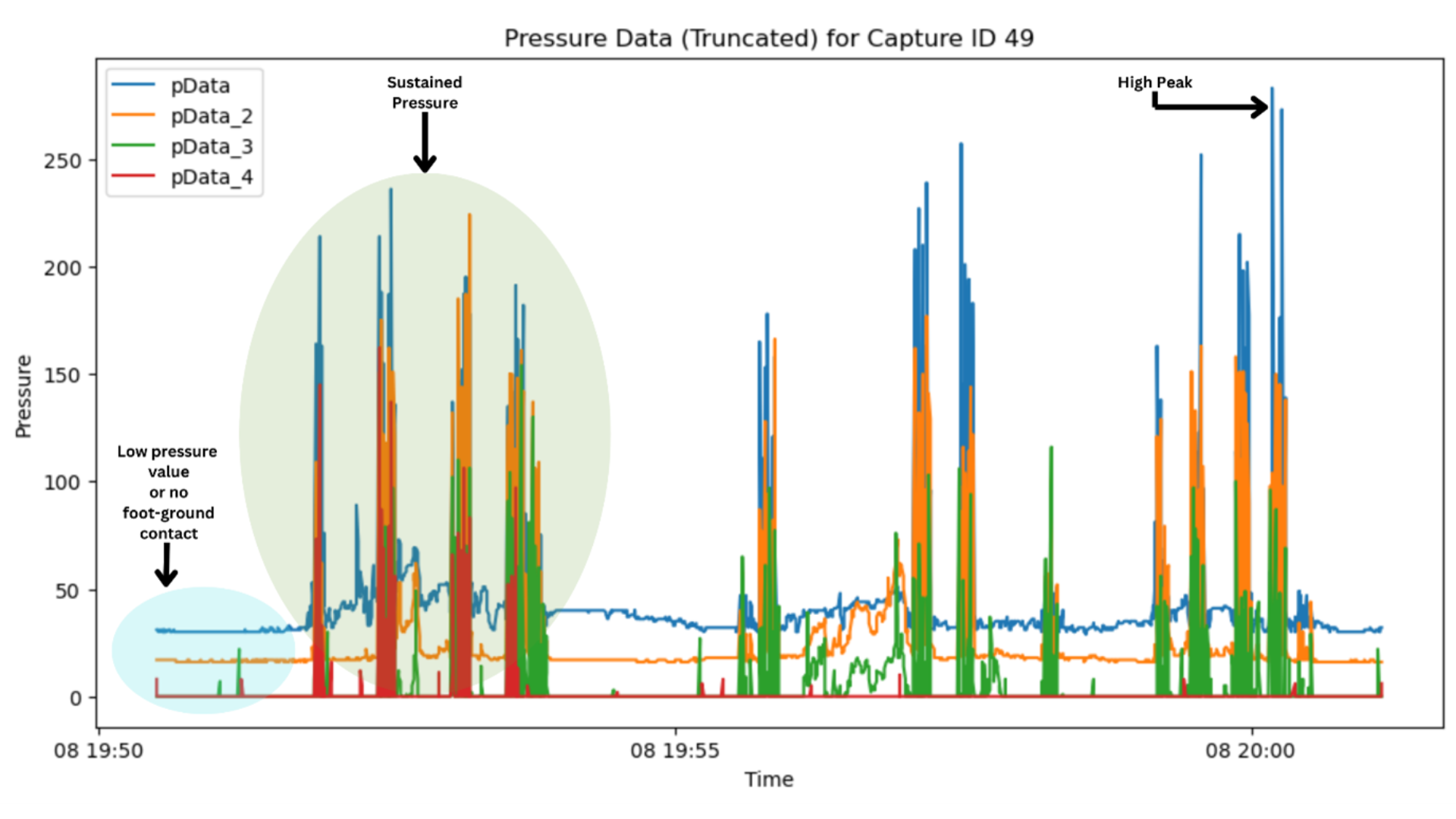}
\caption{Visualization of Selected Patient IDs Based on Pressure Data}
\label{fig:pressure_patient_ids}

\end{center}
\end{figure}

Pressure data visualization reveals foot pressure distribution during
activities like walking and standing.

\paragraph{Features}
\label{features}

Foot pressure data provides insights into load distribution and stress
points, which are key indicators for predicting ulcers. Features were
derived from raw sensor data to capture pressure behavior, including:
maximum pressure, mean pressure, number of steps, step period, and area
under the curve. These pressure-based features help understand foot
behavior under stress and walking, allowing more analysis of ulcer
risks.

To improve model accuracy and efficiency, feature selection was applied
to eliminate redundant and irrelevant features The process involved
selecting numeric pressure data columns, eliminating features with
variance below 0.01, and identifying highly correlated features
(correlation coefficient \textgreater{} 0.95) to address
multicollinearity. Removing these redundant features optimized the
dataset for improved analysis and modeling performance, ensuring only
the most relevant pressure features were retained.

\paragraph{Distribution of features across pData(Pressure Data) types}
\label{distribution-of-features-across-pdatapressure-data-types}
\begin{figure}[htbp]
\begin{center}
\includegraphics[width=\linewidth]{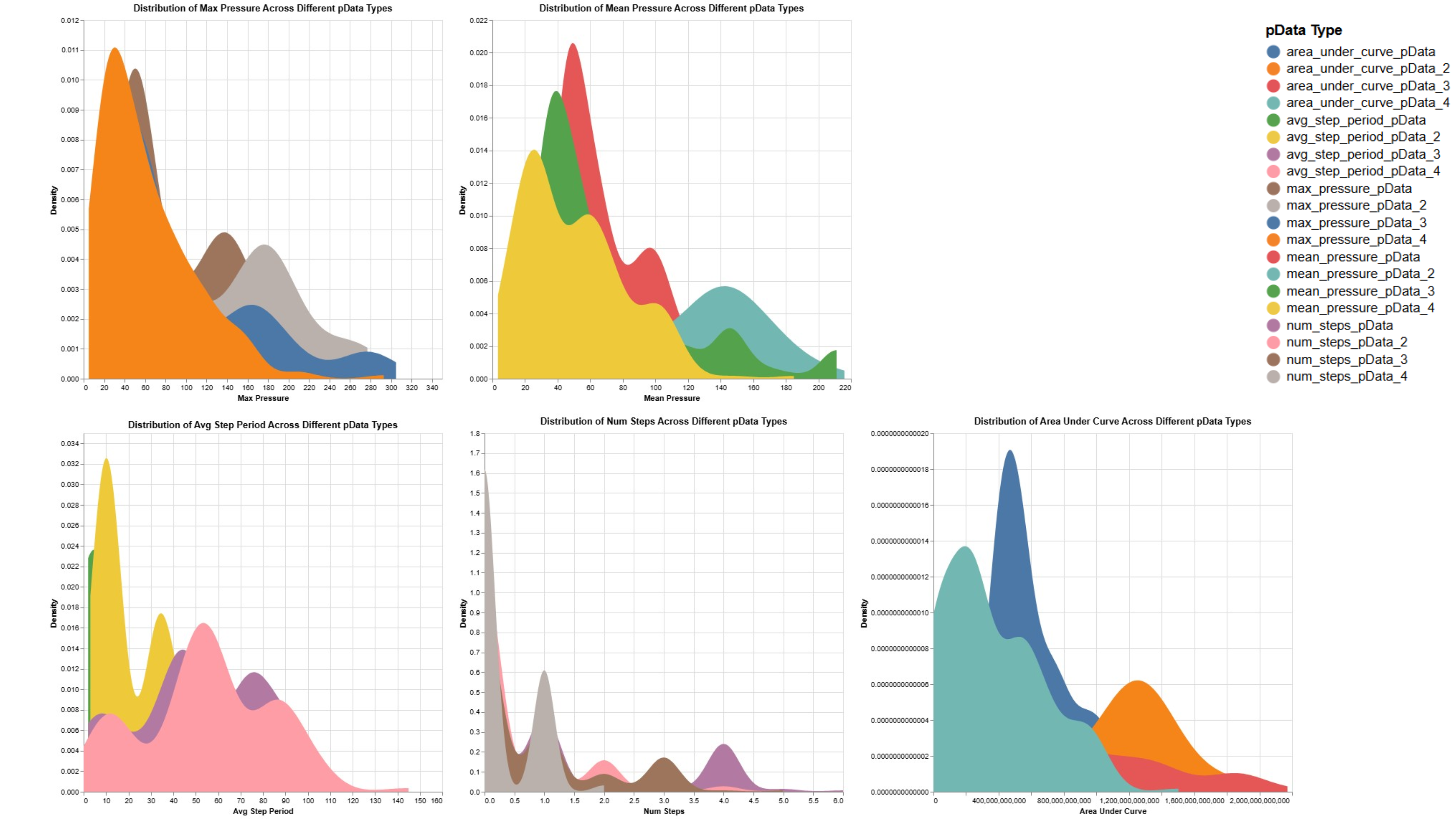}
\caption{Distribution of features across pData(Pressure Data) types}
\label{fig:pressure_features_distribution}

\end{center}
\end{figure}

The analysis revealed that maximum pressure readings were mostly concentrated in lower ranges (0-100 units), indicating generally healthy foot conditions with reduced immediate risk. However, the step analysis showed potential overloading during push-off (shorter step periods in forefoot/toes) and sustained pressure risks (longer step periods in
heel/midfoot). The Area Under Curve (AUC) values were lower, suggesting minimal prolonged pressure exposure, but variations across sensors helped identify areas of sustained pressure risk. These findings provide a basis for future assessment of ulcer risk in diabetic populations, where concentrated or sustained pressure may contribute to tissue breakdown.

\paragraph{Anomaly Detection Using Isolation Forest}
\label{anomaly-detection-using-isolation-forest}

Anomaly detection identifies unusual or rare events in time-series data. Isolation Forest detects subtle anomalies through recursive datasplitting. Future research could explore DBSCAN clustering ortime-series forecasting for long-term trend analysis.

The Isolation Forest algorithm was implemented for anomaly detection in foot pressure data through systematic feature engineering and model optimization, focusing on pressure metrics such as maximum pressure and step counts. Through extensive hyperparameter tuning using Grid Search, the optimal configuration was determined to be 100 trees, max\_samples
of 0.6, and a contamination rate of 0.05. A contamination rate of 0.05 was selected, a common baseline for anomaly detection in health monitoring datasets. The anomaly score for a data point $x$ is formally 
defined as:
\begin{equation}
s(x, n) = 2^{-\frac{E[h(x)]}{c(n)}}
\label{eq:ifscore}
\end{equation}
where $E[h(x)]$ is the average path length of $x$ across all trees in the 
ensemble, and $c(n) = 2H(n-1) - \frac{2(n-1)}{n}$ is the expected path length 
for a dataset of size $n$, with $H(i)$ denoting the harmonic number. Scores 
close to 1 indicate anomalies, scores near 0.5 indicate normal points, and 
scores significantly below 0.5 indicate normal behaviour. In the scikit-learn 
implementation used here, the \texttt{score\_samples()} function returns the 
negated raw scores, so values below zero correspond to anomalous points. The 
model generated a binary classification (1 for normal, $-$1 for anomalous) 
by applying the contamination threshold of 0.05, which labels the bottom 5\% of scores as anomalous.

\paragraph{Feature Importance for Pressure Anomaly Detection for Isolation Forest}
\label{feature-importance-for-pressure-anomaly-detection-for-isolation-forest}
\begin{figure}[htbp]
\begin{center}
\includegraphics[width=\linewidth]{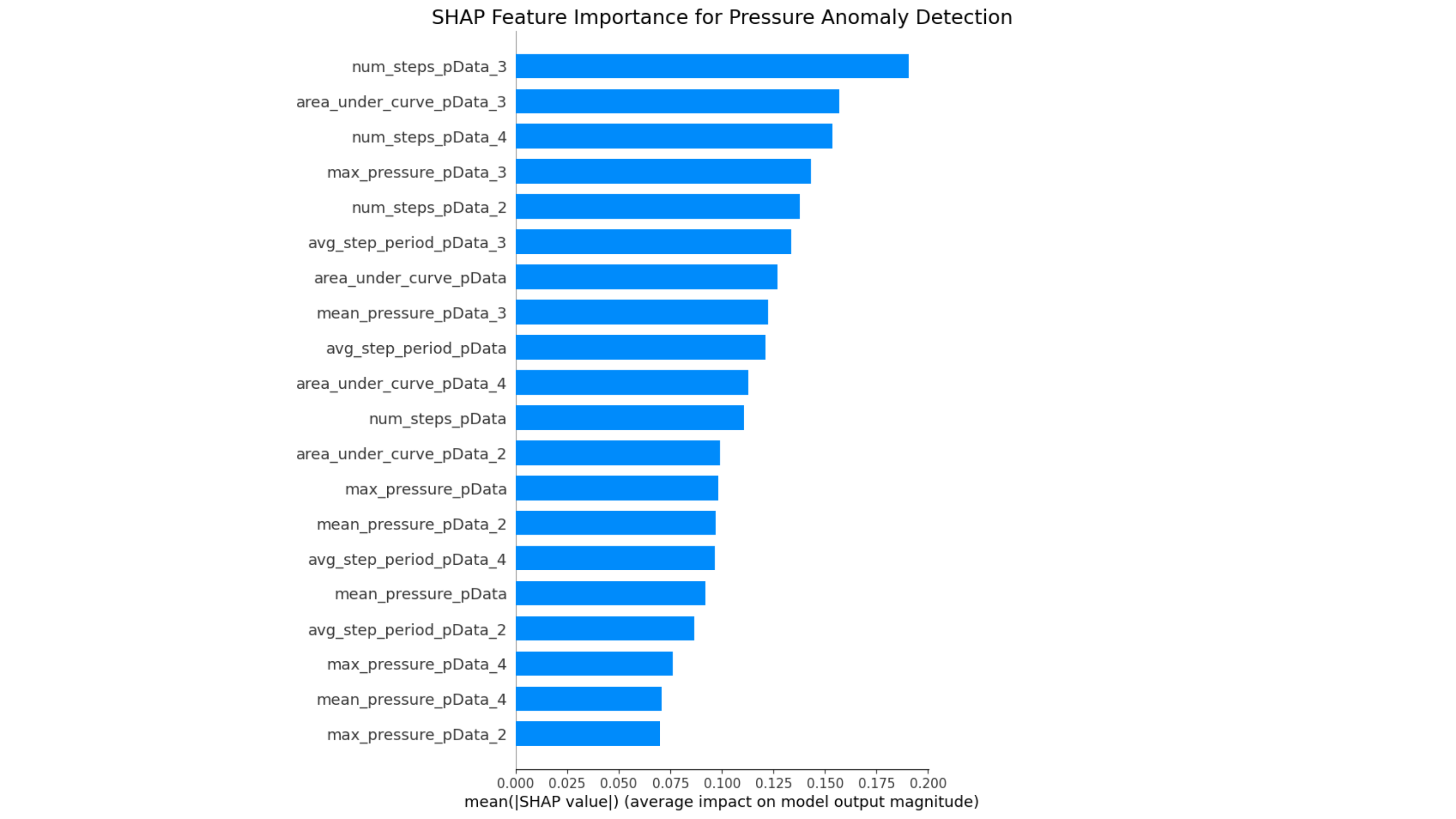}
\caption{Feature Importance for Pressure Anomaly Detection for Isolation Forest}
\label{fig:pressure_feature_importance}

\end{center}
\end{figure}

The most crucial features for detecting anomalies in pressure data are
the number of steps and area under the pressure curve from sensor 3, as
well as the number of steps from sensor 4. Maximum pressure from sensor
3, average step period from sensor 3, and number of steps from sensor 2
also play moderately important roles. In contrast, maximum and mean
pressures from sensors 2 and 4 are less influential. The high importance
of features from sensor 3, which may correspond to the midfoot or heel
area, suggests that changes in gait mechanics or sustained pressure in
this region are particularly indicative of ulcer risk. For instance, an
increased number of steps or a larger area under the curve could signal
compensatory gait patterns or prolonged stress that precedes tissue
breakdown.

\paragraph{Visualization of Isolation Forest Results}

The visualizations depict anomalies detected by the Isolation Forest as
red points, while normal readings are green.
\begin{figure}[htbp]
\begin{center}
\includegraphics[width=\linewidth]{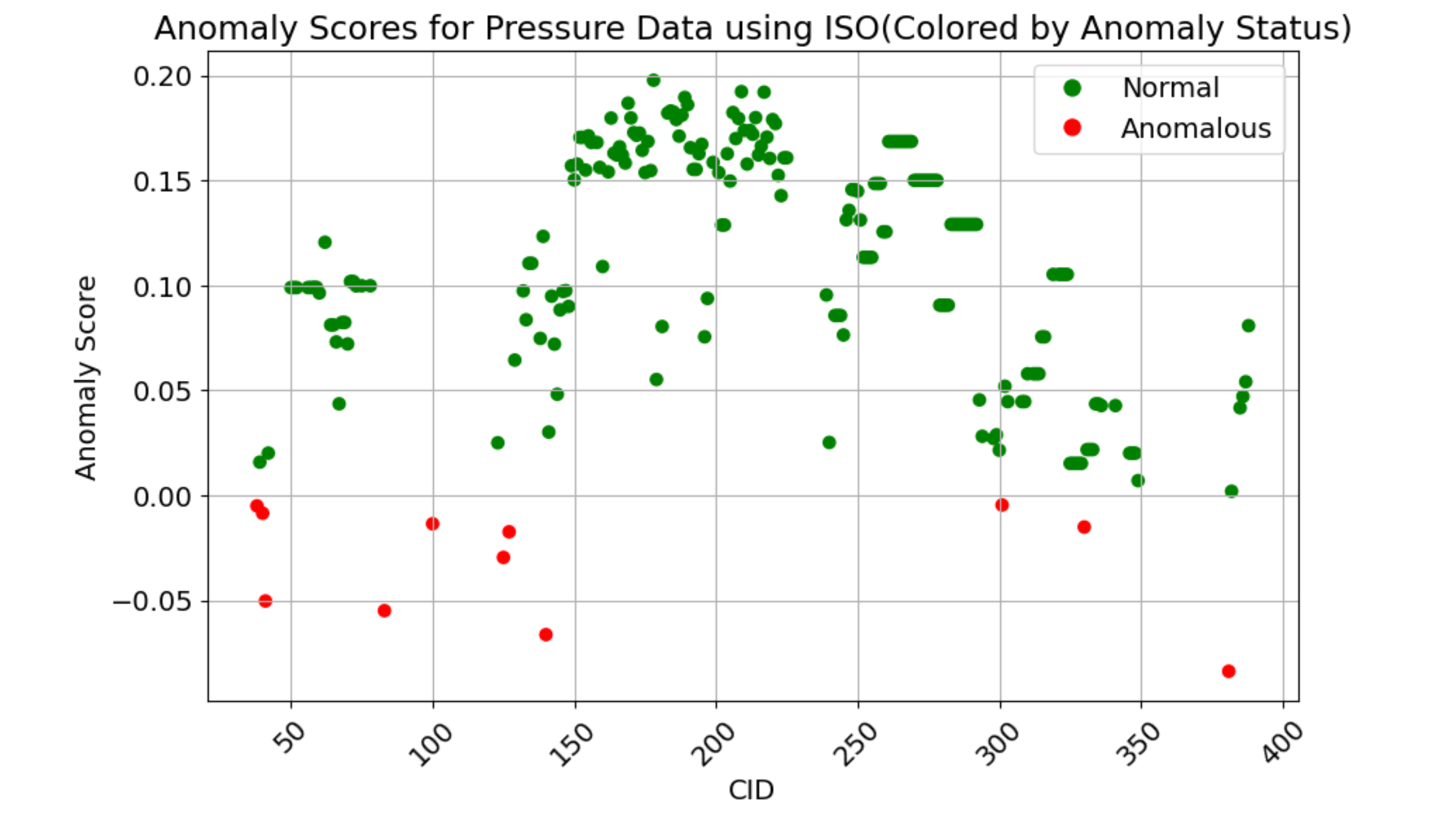}
\caption{Pressure Data (Anomaly Scores for Pressure Data)}
\label{fig:pressureanomalyscores}

\end{center}
\end{figure}

Pressure Data (Anomaly Scores for Pressure Data):Positive scores denote
normal behavior predominates. However, negative dips around Patient ID
140 and 380, falling below -0.07, indicate potential anomalous pressure
data points requiring investigation.

\paragraph{Concentration of Anomalies for some most important features}
\begin{figure}[htbp]
\begin{center}
\includegraphics[width=\linewidth]{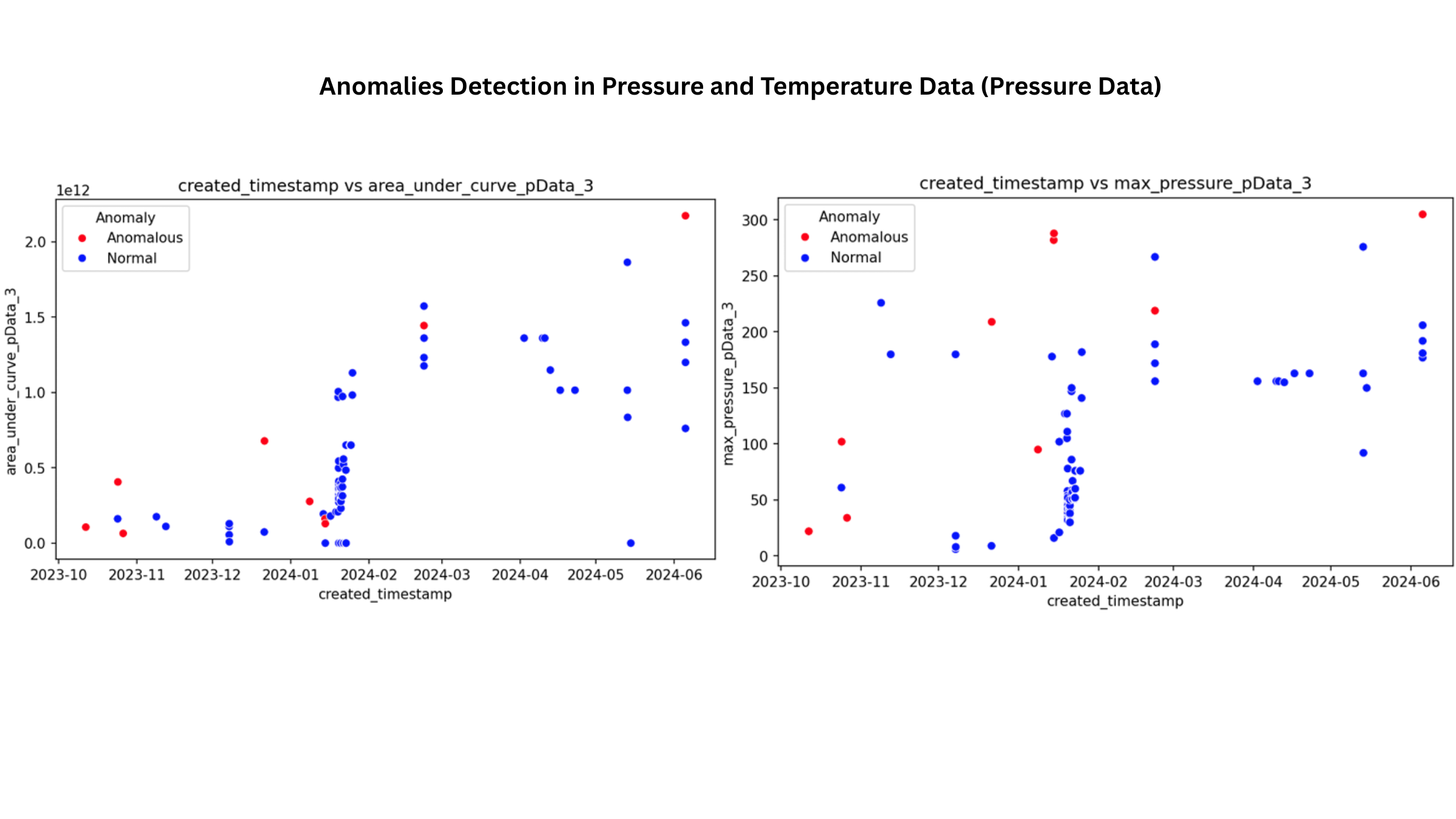}
\caption{Concentration of Anomalies for some most important features (Pressure - ISO)}
\label{fig:pressureanomalyconcentration}

\end{center}
\end{figure}

The anomalies for pressure data appear over a wide range of values, which means 
that the system experienced both very low-pressure events and extremely high-pressure 
events, particularly around October 2023, January and June 2024 (Figure~\ref{fig:pressureanomalyconcentration}). 
This is further supported by the plots showing max pressure and area under the 
curve, where large deviations from normal conditions are more evident.

\begin{figure}[htbp]
\begin{center}
\includegraphics[width=\linewidth]{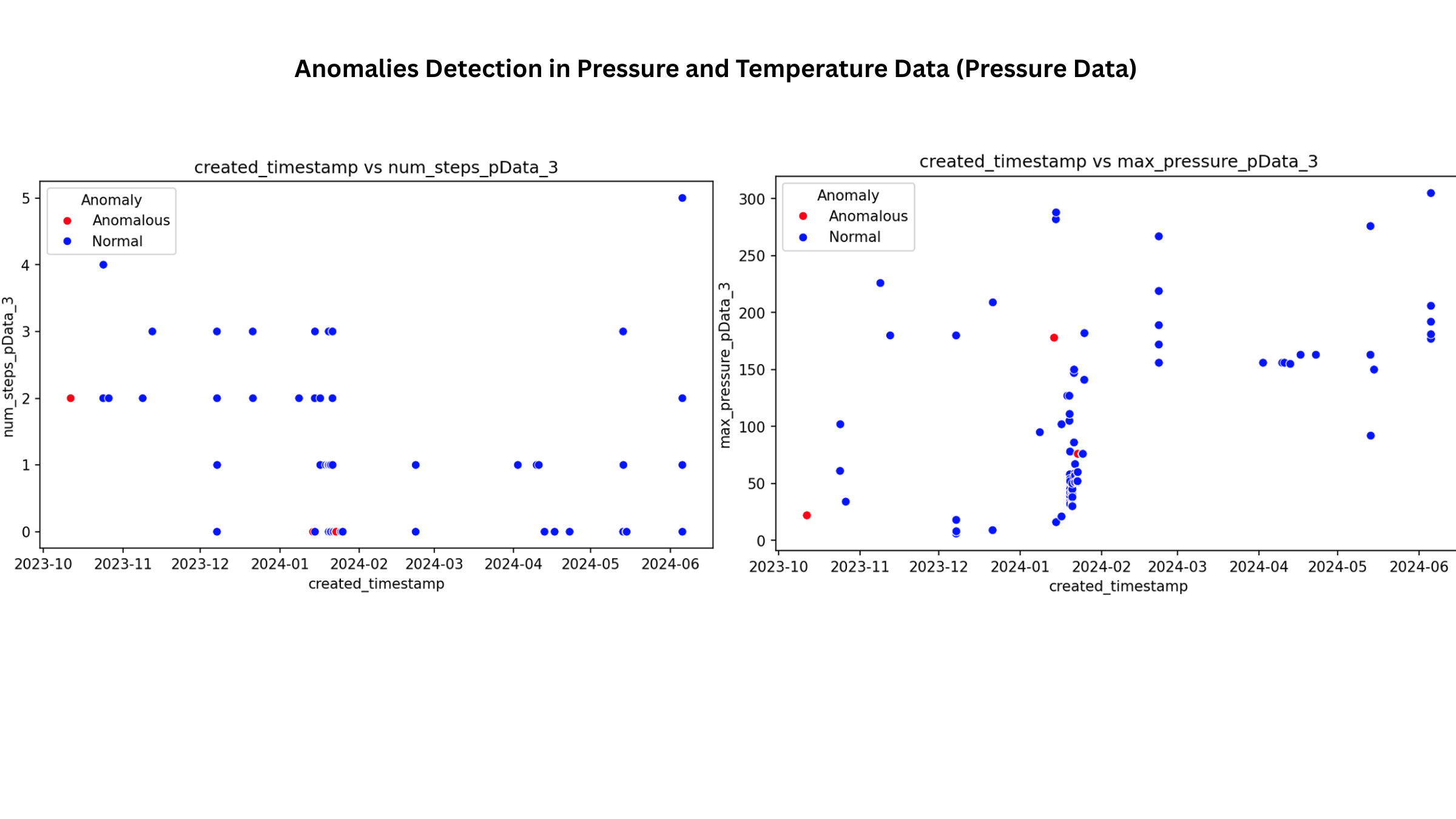}
\caption{Concentration of Anomalies for some most important features (Pressure - KNN)}
\label{fig:pressureknnanomalyconcentration}

\end{center}
\end{figure}

It should be noted that the anomalous sessions 
identified in Figure~\ref{fig:pressureanomalyconcentration} (Isolation Forest) 
differ from those shown in Figure~\ref{fig:pressureknnanomalyconcentration} 
(KNN/LOF). This is a direct consequence of the two algorithms using fundamentally 
different detection mechanisms. Isolation Forest identifies anomalies by measuring 
how easily a point can be isolated through random partitioning, making it sensitive 
to broadly distributed, low-density deviations across the feature space. KNN/LOF, 
on the other hand, flags points whose local neighbourhood density is significantly 
lower than that of their neighbours, making it more sensitive to sharp, localised 
extremes. The two figures therefore reflect complementary perspectives on the same 
data rather than contradictory results.

\paragraph{K-Nearest Neighbors (KNN) for Anomaly Detection}
\label{k-nearest-neighbors-knn-for-anomaly-detection}

The K-Nearest Neighbors (KNN) Anomaly Detection model, specifically the
Local Outlier Factor (LOF) variation, was applied to pressure data to
identify anomalies by calculating the distance of each data point to its
nearest neighbors, considering both distance and relative density of
neighboring points. The model was configured with k = 20 neighbors and a
contamination rate of 0.05. The value for k was determined experimentally to best capture local data density, and the contamination parameter was set to 0.05 to align with the Isolation Forest model. Relevant pressure features were selected and standardized using
StandardScaler, and the LOF model assigned each data point an anomaly 
score defined as:
\begin{equation}
\text{LOF}_k(p) = \frac{1}{|N_k(p)|} \sum_{o \in N_k(p)} 
\frac{\text{lrd}_k(o)}{\text{lrd}_k(p)}
\label{eq:lofscore}
\end{equation}
where $N_k(p)$ is the set of $k$ nearest neighbours of point $p$, and 
$\text{lrd}_k(p)$ is the local reachability density, defined as the inverse 
of the average reachability distance of $p$ with respect to its neighbours. 
A score close to 1 indicates that $p$ has a similar density to its neighbours 
and is considered normal. Scores significantly greater than 1 indicate that 
$p$ lies in a region of much lower density than its neighbours and is flagged 
as anomalous. As with Isolation Forest, the binary label (Normal or Anomalous) 
was derived by applying the contamination threshold of 0.05, labelling the top 
5 percent of LOF scores as anomalous.

\begin{figure}[htbp]
\begin{center}
\includegraphics[width=\linewidth]{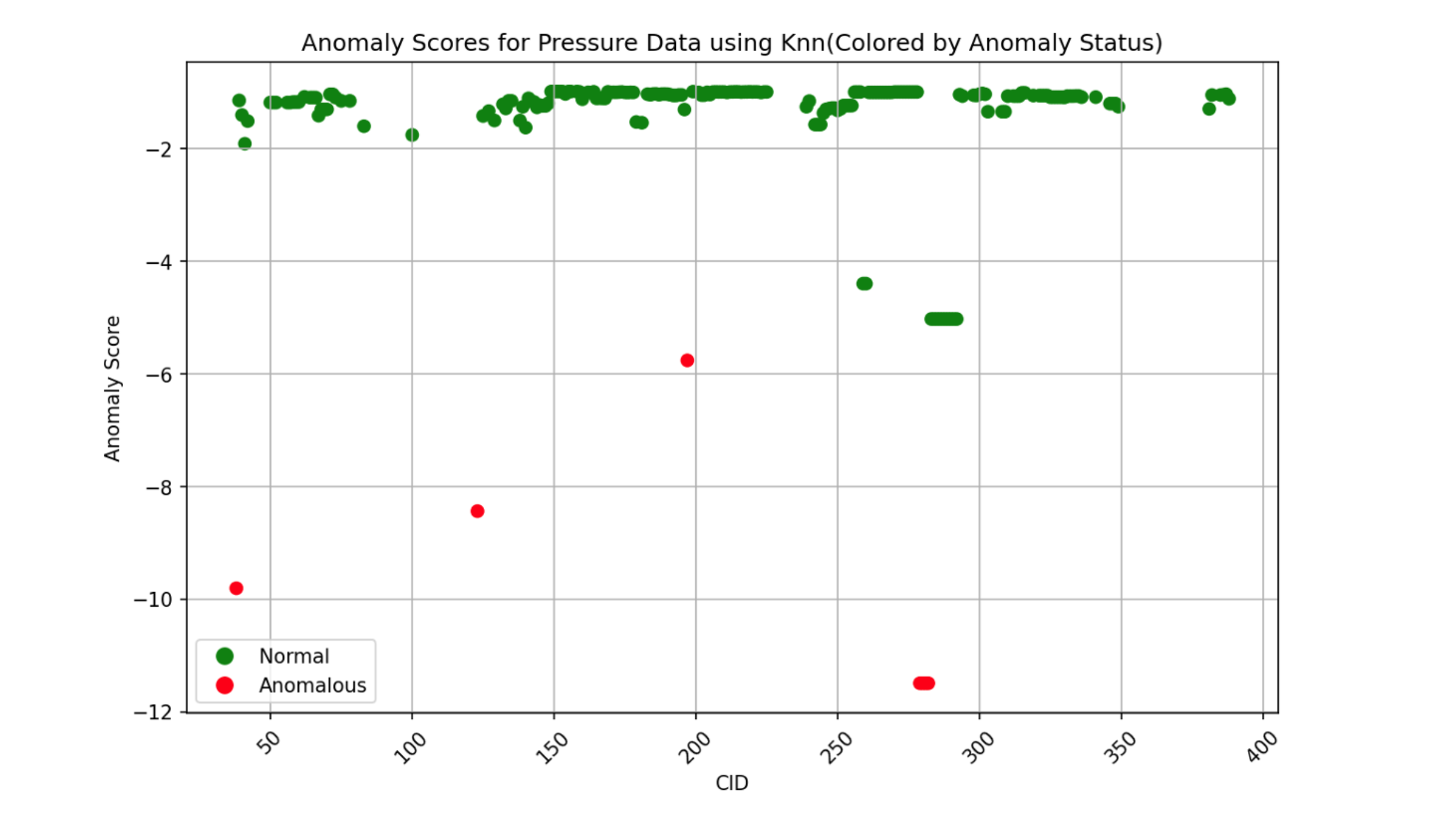}
\caption{KNN Anomaly Detection Results (Pressure Data)}
\label{fig:pressure_knn_results}

\end{center}
\end{figure}

\paragraph{Concentration of Anomalies for some most important features}

Pressure data anomalies emerged around October 2023 and January 2024,
with readings deviating significantly from expected trends, indicating
abnormal pressure fluctuations and potential unusual behavior or
environmental disturbances during those periods.

\subsubsection{3.2 Temperature-Based Analysis}
\label{temperature-based-analysis}

\paragraph{Visualization of Selected Capture IDs Based on Temperature Data}
\label{visualization-of-selected-capture-ids-based-on-temperature-data}
\begin{figure}[htbp]
\begin{center}
\includegraphics[width=\linewidth]{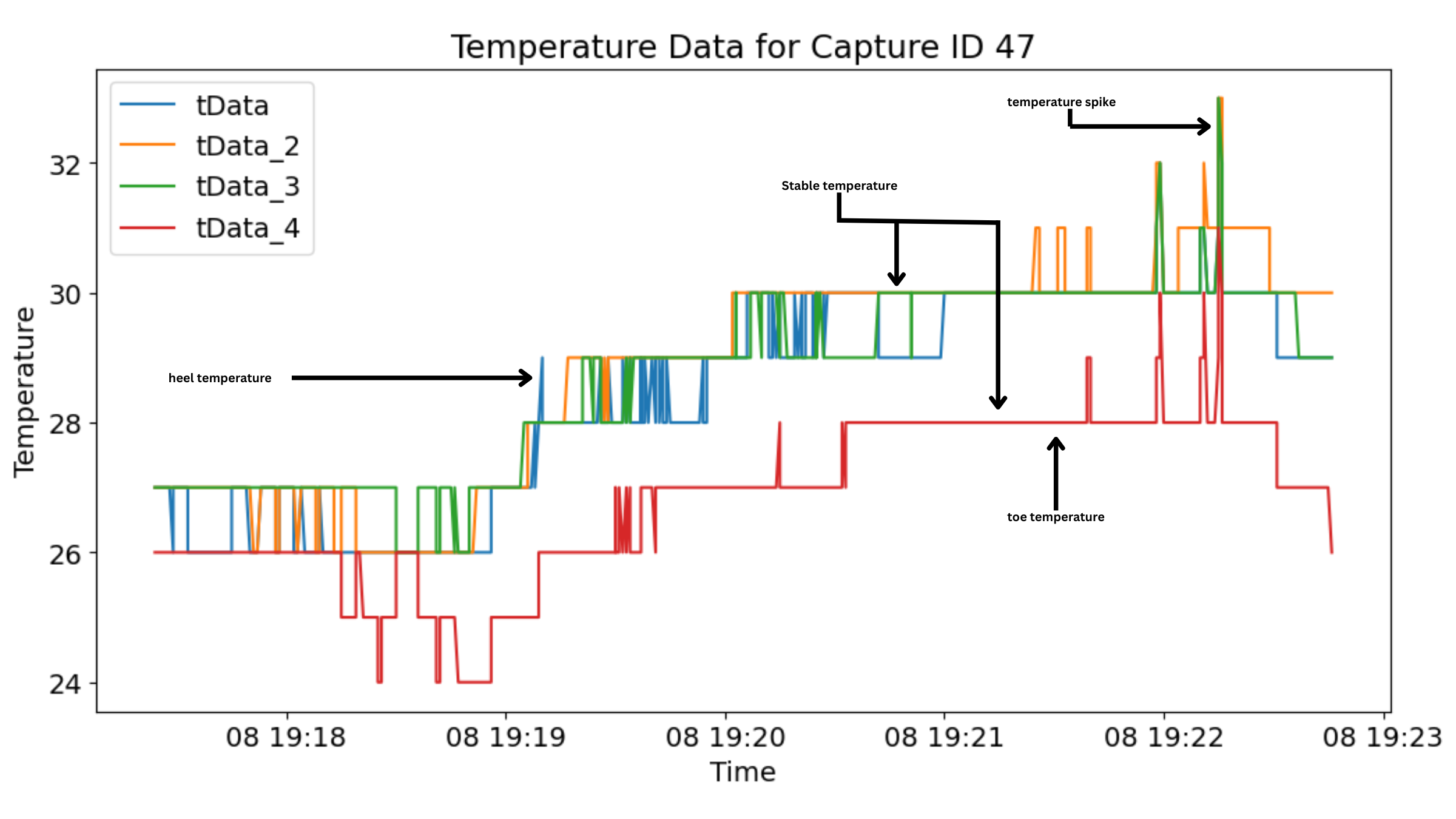}
\caption{Visualization of Selected Capture IDs Based on Temperature Data}
\label{fig:temperature_capture_ids}

\end{center}
\end{figure}

Temperature data visualization reveals temperature patterns and thermal
distribution across foot regions.

\paragraph{Features}
\label{features-1}

Temperature data provides potentially important signals for future foot 
ulcer risk research, as abnormal patterns in clinical populations may be 
associated with inflammation or poor circulation — though this relationship 
cannot be confirmed in the present healthy-subject dataset.

Key features extracted from temperature data include Maximum, minimum,
and mean temperatures, Temperature change/variation rates.

These features capture temporal and physical foot health aspects,
improving predictive analytics for ulcer risk identification.

\paragraph{Distribution of features across tData(Temperature data) types}
\label{distribution-of-features-across-tdatatemperature-data-types}
\begin{figure}[htbp]
\begin{center}
\includegraphics[width=\linewidth]{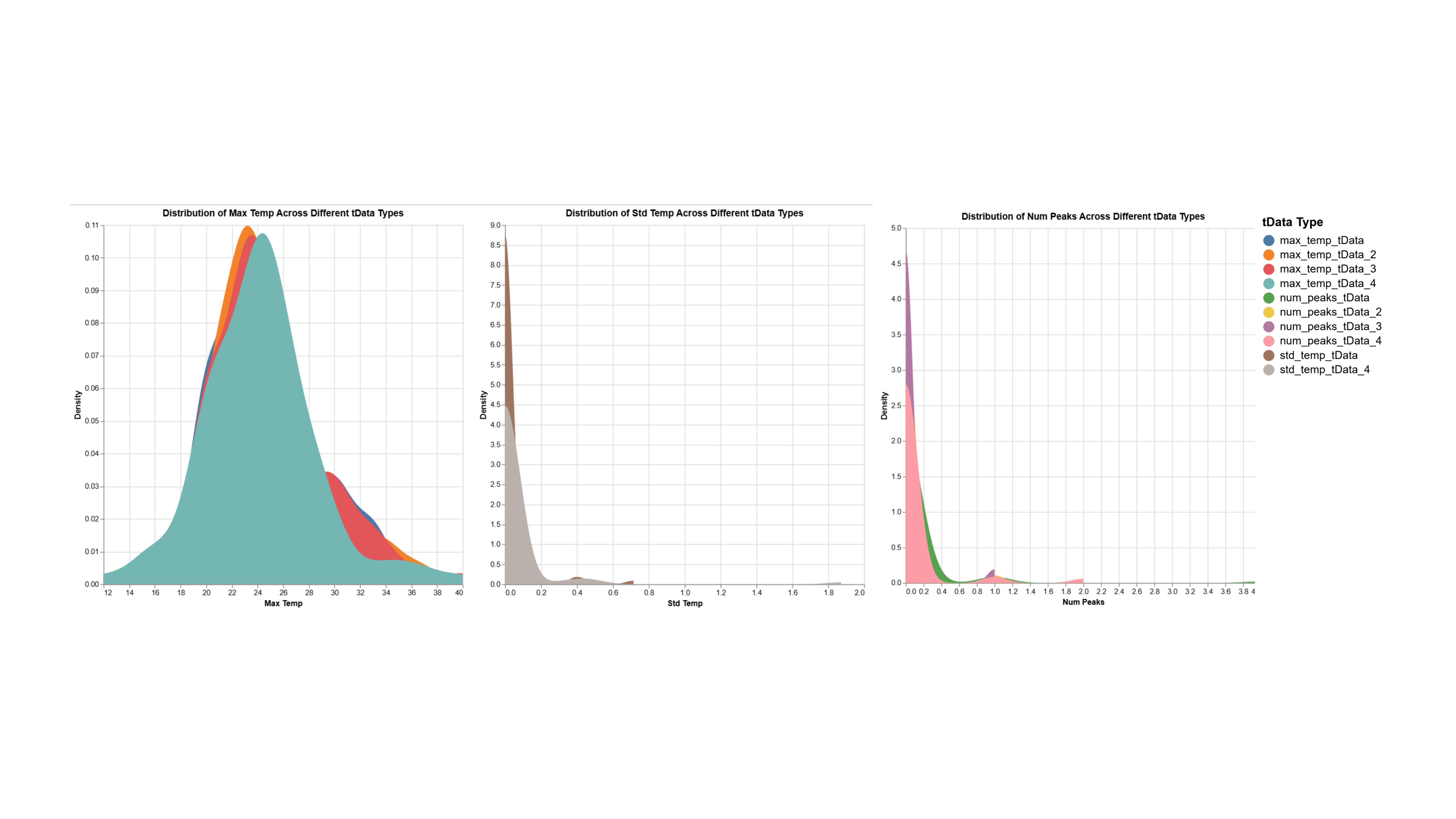}
\caption{Distribution of features across tData(Temperature data) types}
\label{fig:temperature_features_distribution}

\end{center}
\end{figure}

The analysis of foot temperature metrics revealed important insights
into foot health assessment. The maximum temperatures measured fell
within the typical physiological range of 20-26°C, although variations
were observed among sensors, potentially due to placement or performance
differences. Some sensors detected higher maximum temperatures, which
could indicate areas at risk of inflammation or ulcers. Overall, the low
temperature standard deviation and minimal peaks suggested stable foot
temperatures during measurements, generally indicating favorable foot
health. However, it is crucial to continuously monitor foot
temperatures, particularly in diabetic patients, as consistent
temperature elevations could mask underlying issues. Temperature
variations can signal inflammation or potential complications,
highlighting the importance of ongoing monitoring despite apparent
temperature stability.

\paragraph{Anomaly Detection Using Isolation Forest}
\label{anomaly-detection-using-isolation-forest-1}

The Isolation Forest algorithm was applied to temperature data using the same methodology and hyperparameter configuration as the pressure analysis. The model produced a binary classification system (1 for
normal, -1 for anomalous) based on anomaly scores, where lower scores
indicated a higher likelihood of anomalous foot temperature patterns,
such as inflammation or hotspots.

\paragraph{Feature Importance for Temperature Anomaly Detection for Isolation Forest}
\label{feature-importance-for-temperature-anomaly-detection-for-isolation-forest}
\begin{figure}[htbp]
\begin{center}
\includegraphics[width=\linewidth]{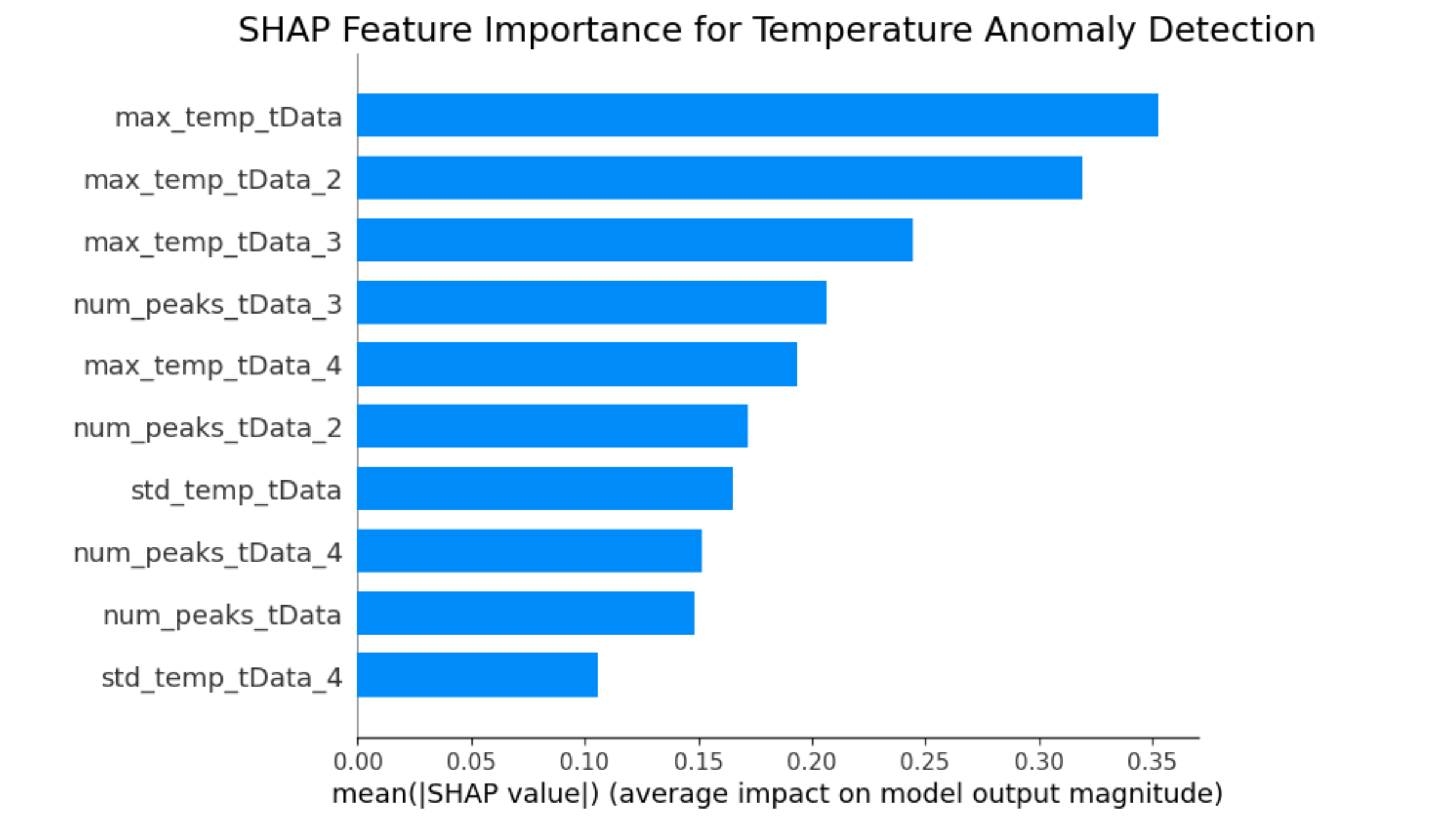}
\caption{Feature Importance for Temperature Anomaly Detection for Isolation Forest}
\label{fig:temperature_feature_importance}

\end{center}
\end{figure}

For temperature data, the most influential features are maximum
temperature features (max\_temp\_tData, max\_temp\_tData\_2,
max\_temp\_tData\_3), with sudden spikes or fluctuations in maximum
temperatures being the strongest signals of abnormal behavior.
Temperature peaks (num\_peaks\_tData\_3) also play a moderately
important role in detecting anomalies, while temperature standard
deviation (std\_temp\_tData\_4) has a lesser influence on the model's
decisions, collectively helping to identify abnormal temperature
patterns such as sudden spikes, fluctuations, and peaks that may
indicate inflammation, hotspots, or poor temperature regulation.

\paragraph{Visualization of Isolation Forest Results}
\begin{figure}[htbp]
\begin{center}
\includegraphics[width=\linewidth]{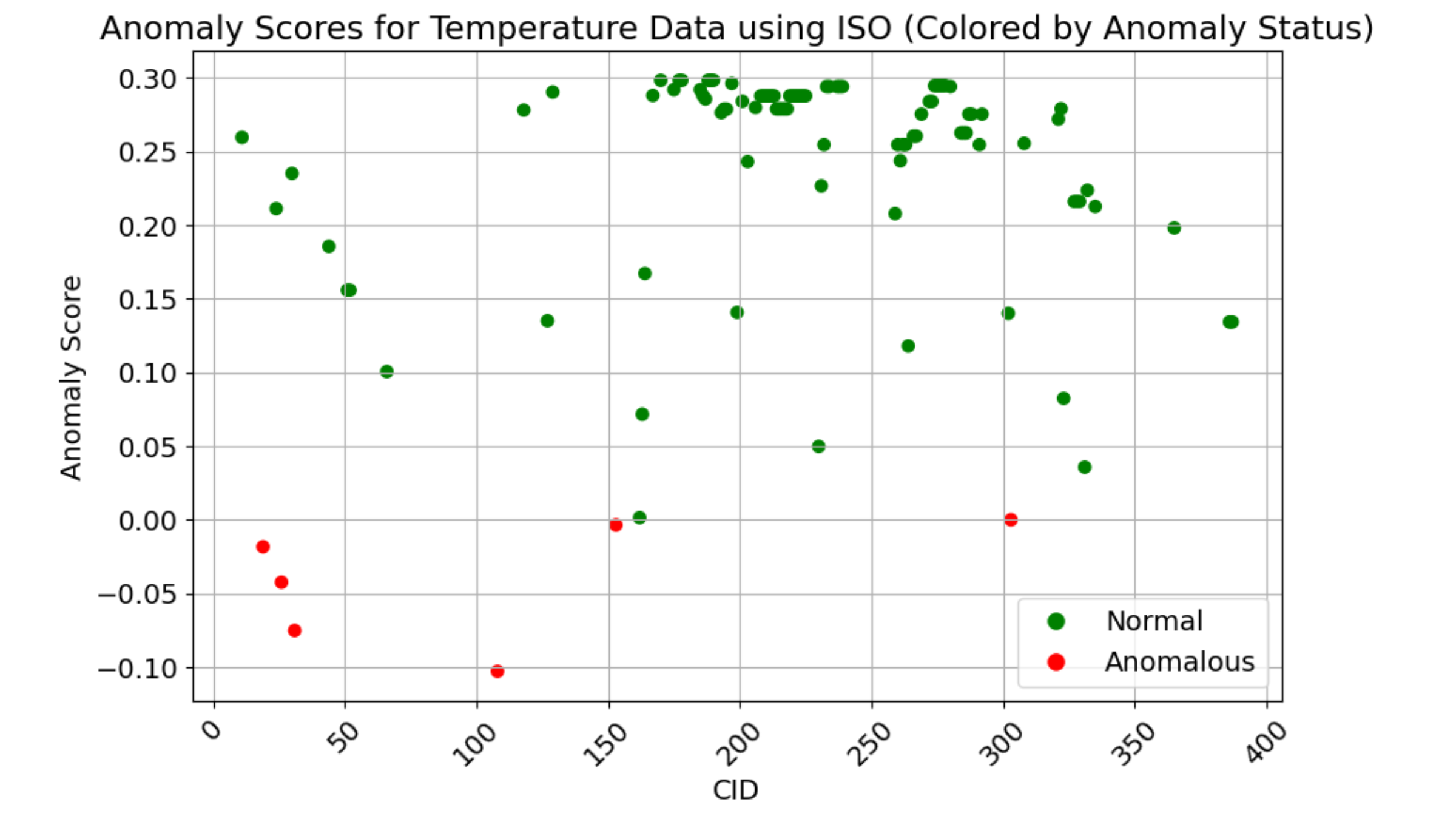}
\caption{Temperature Data (Anomaly Scores for Temperature Data)}
\label{fig:temperature_anomaly_scores}

\end{center}
\end{figure}

Temperature Data (Anomaly Scores for Temperature Data):The anomaly
scores predominantly range from 0.3 to 0.2, signifying normal behavior.
However, distinct drops below zero are observed at several points,
indicating anomalous behavior. The most significant deviations from the
norm occur around Patient ID's 0-50, 110, 150, and 300, as evidenced by
the sharp temperature reading anomalies.

\paragraph{Concentration of Anomalies for some most important features}
\begin{figure}[htbp]
\begin{center}
\includegraphics[width=\linewidth]{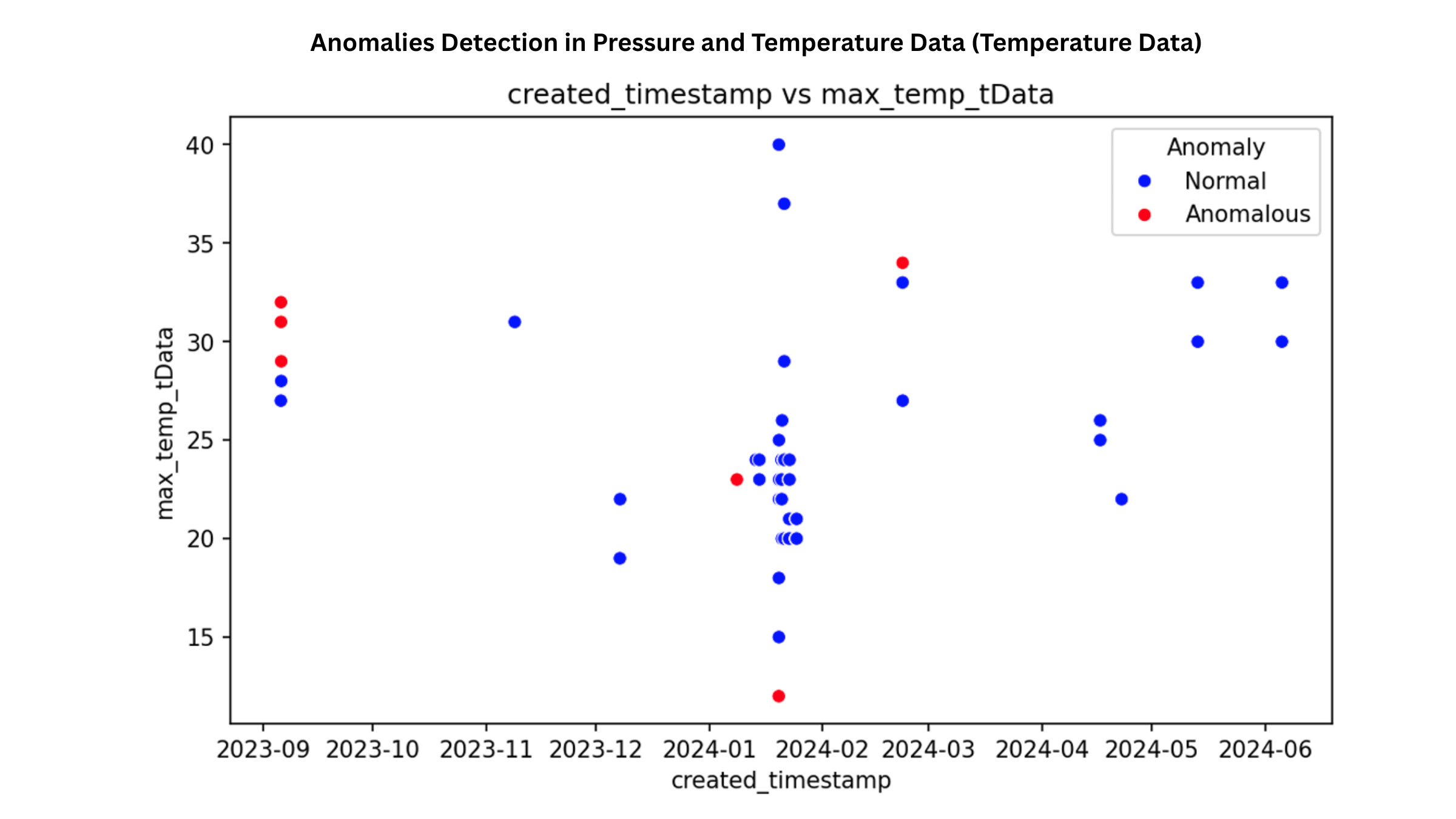}
\caption{Concentration of Anomalies for some most important features (Temperature - ISO)}
\label{fig:temperatureanomalyconcentration}

\end{center}
\end{figure}

The temperature data exhibits anomalies during September 2023, January
and February 2024, characterized by unusually high or low temperatures
deviating from typical values. Most anomalies occur around typical
temperature ranges (e.g., 20°C to 35°C), but some outliers exist,
particularly during early 2024 with significantly higher or lower
temperatures than expected.

\paragraph{K-Nearest Neighbors (KNN) for Anomaly Detection}
\label{k-nearest-neighbors-knn-for-anomaly-detection-1}

The KNN/LOF model was also applied to temperature data, leveraging its
ability to detect anomalies based on distance and density. With the same
configuration as the pressure model (k = 20 and contamination rate of
0.05), the model identified unusual temperature patterns, labeling data
points as 'Normal' or 'Anomalous' and providing anomaly scores, where
lower scores indicated a higher likelihood of temperature-related
anomalies, such as sudden spikes or fluctuations that could signal
potential health issues.
\begin{figure}[htbp]
\begin{center}
\includegraphics[width=\linewidth]{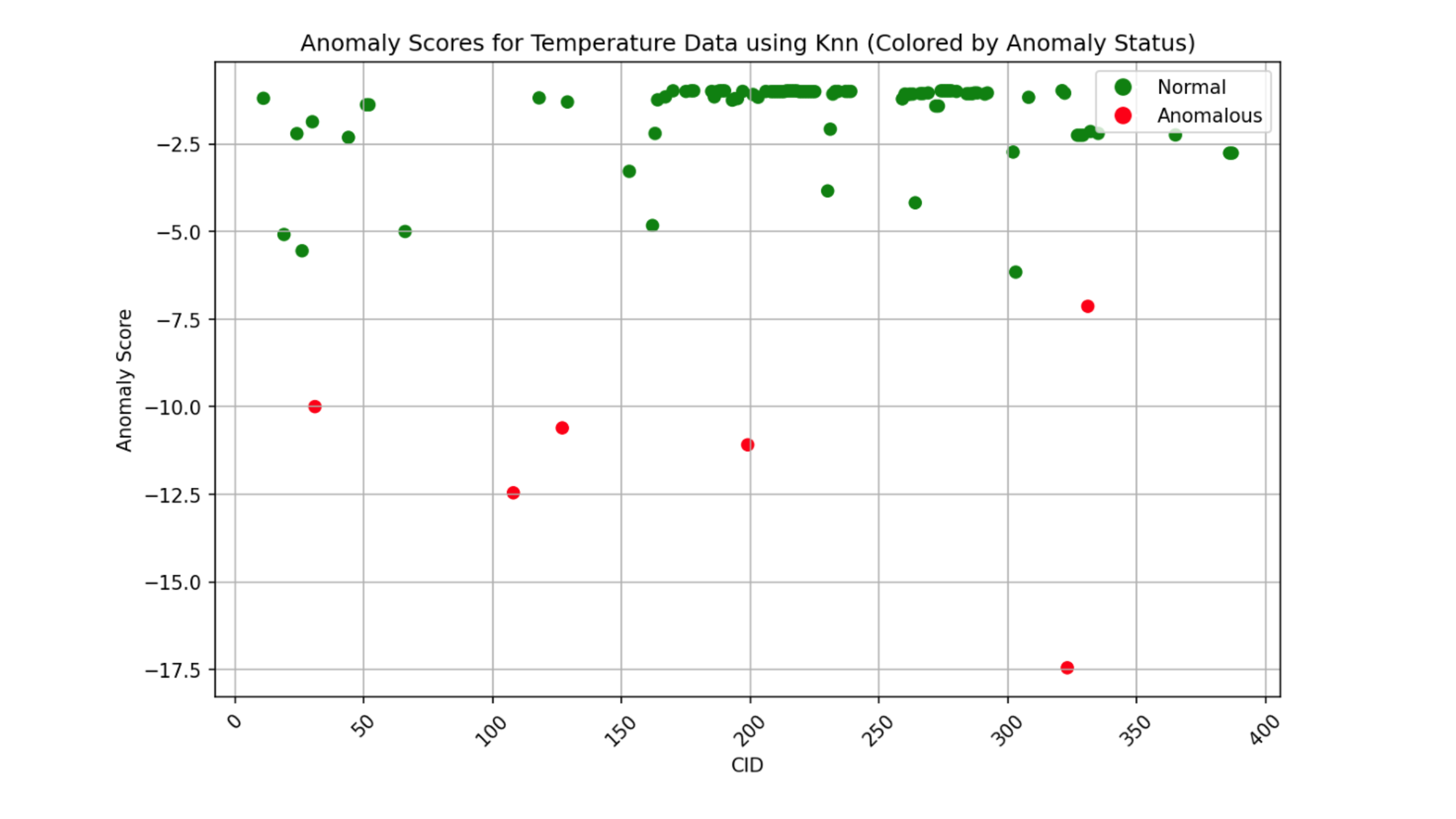}
\caption{KNN Anomaly Detection Results (Temperature Data)}
\label{fig:temperature_knn_results}

\end{center}
\end{figure}

\paragraph{Concentration of Anomalies for some most important features}
\begin{figure}[htbp]
\begin{center}
\includegraphics[width=\linewidth]{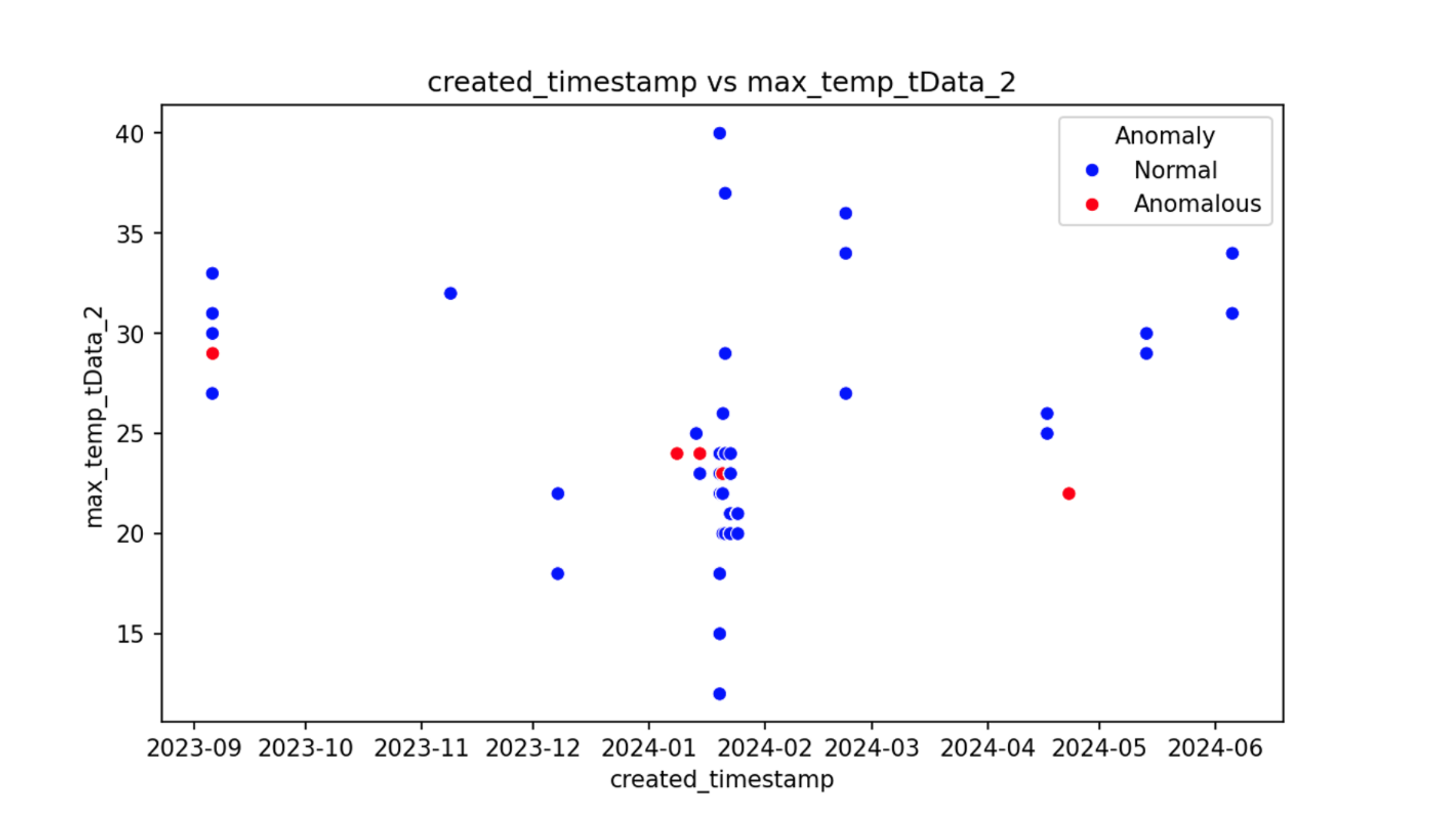}
\caption{Concentration of Anomalies for some most important features (Temperature - KNN)}
\label{fig:temperatureknnanomalyconcentration}

\end{center}
\end{figure}

Temperature data anomalies occurred across a wide range, 
notably around September 2023, January and April 2024 (Figure~\ref{fig:temperatureknnanomalyconcentration}), 
indicating unusually high and low temperatures. While most readings were typical, 
some points deviated significantly, suggesting potential issues or unusual 
environmental conditions during those periods.

Both temperature and pressure data exhibited concentrated 
anomalies from September to October 2023, January 2024, and April 2024 
(Figures~\ref{fig:pressureanomalyconcentration}--\ref{fig:temperatureknnanomalyconcentration}). 
However, these raw counts may partly reflect varying sampling density across 
months. To address this, anomaly rates were normalised by the number of capture 
sessions per month (Table~\ref{tab:normalizedanomalies}). For pressure data 
using Isolation Forest, January 2024 showed a normalised rate of 2.7\%, February 
14.3\%, and June 16.7\%, confirming elevated anomaly prevalence beyond sampling 
density alone. This normalisation strengthens the evidence for temporal clustering 
of statistical deviations.

\begin{table}[htbp]
\centering
\caption{Normalised anomaly rates (Isolation Forest, pressure data) by month, 
as \% of capture sessions flagged anomalous. September 2023 data are excluded 
due to limited capture sessions in the first month of data collection (September 3--30, 2023).}
\label{tab:normalizedanomalies}
\begin{tabular}{lc}
\toprule
Month & Anomaly Rate (\%) \\
\midrule
Oct 2023 & 75.0 \\
Nov 2023 & 0.0 \\
Dec 2023 & 4.8 \\
Jan 2024 & 2.7 \\
Feb 2024 & 14.3 \\
Apr 2024 & 0.0 \\
May 2024 & 5.6 \\
Jun 2024 & 16.7 \\
\bottomrule
\end{tabular}
\end{table}

\subsection{Combined Visualization Analysis}
\label{combined-visualization-analysis}

\paragraph{Heatmaps of Pressure Metrics by Month}
\label{heatmaps-of-pressure-metrics-by-month}
\begin{figure}[htbp]
\begin{center}
\includegraphics[width=\linewidth]{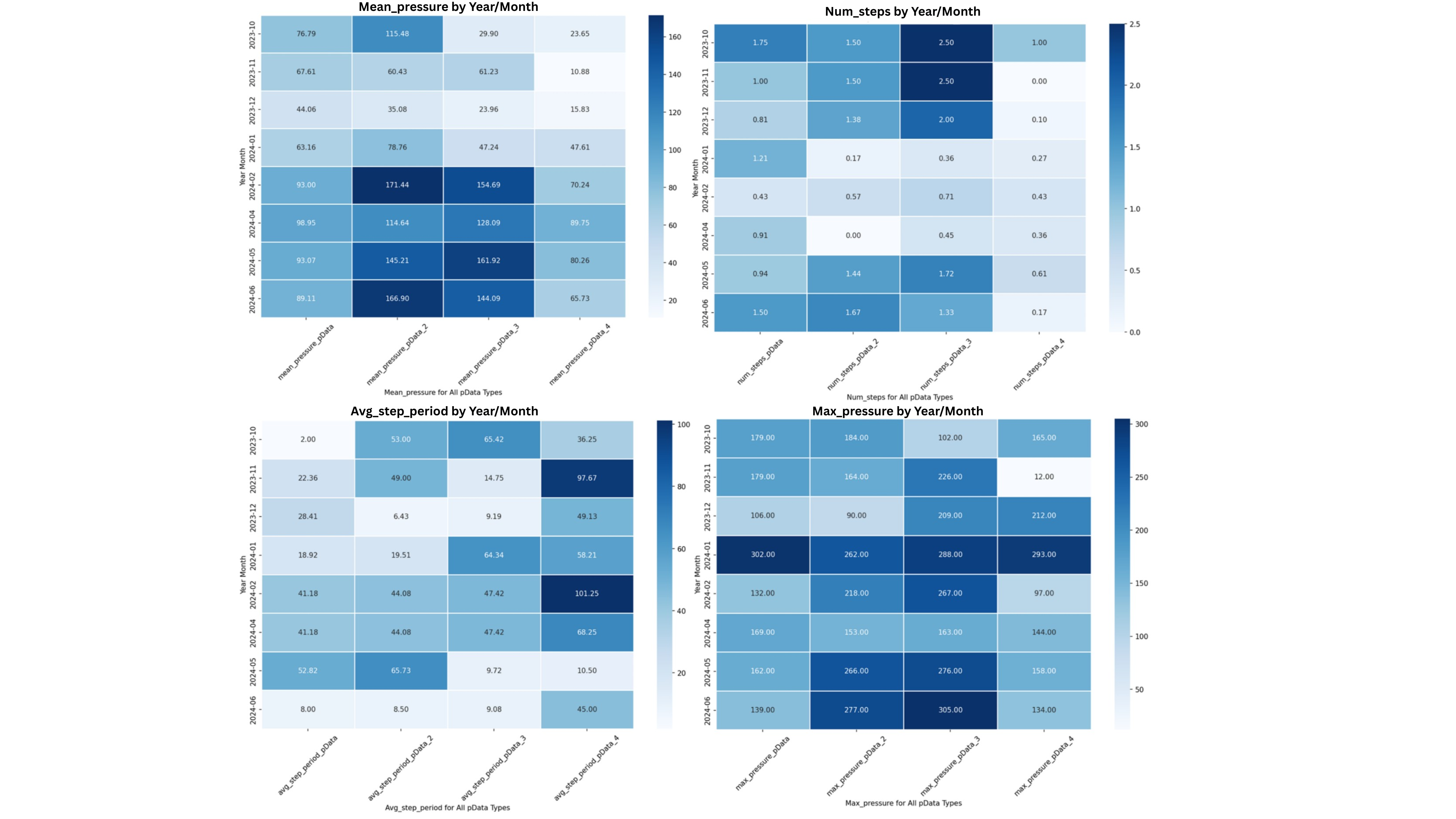}
\caption{Heatmaps of Pressure Metrics by Month (max pressure, mean pressure, 
number of steps, and average step period across four sensors, October 2023 -- 
June 2024). January 2024 shows peak maximum pressures across all sensors, 
while February 2024 exhibits the longest average step periods, indicating 
potential gait irregularities.}

\label{fig:pressure_heatmaps}

\end{center}
\end{figure}

\par\null

Heatmap analysis of pressure metrics from October 2023 to June 2024 showed significant monthly variations. Maximum pressure peaked in January and June 2024, while extended step periods indicated potential risk in November 2023 and February 2024. The ISO algorithm detected a broad range of anomalies, whereas KNN focused on flagging extreme low-pressure events.

\paragraph{Heatmaps of Temperature Metrics by Month}
\label{heatmaps-of-temperature-metrics-by-month}
\begin{figure}[htbp]
\begin{center}
\includegraphics[width=\linewidth]{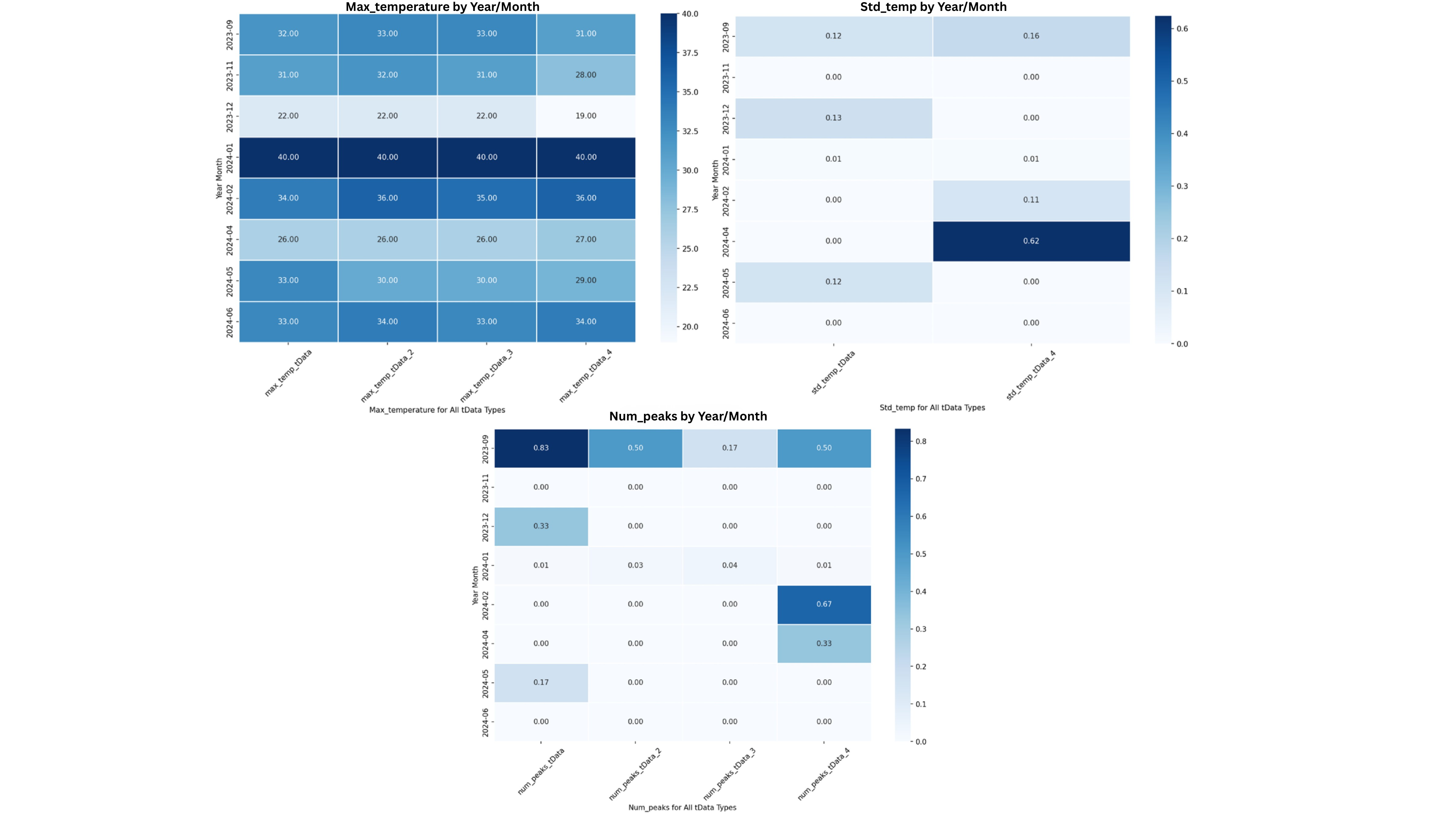}
\caption{Heatmaps of Temperature Metrics by Month}
\label{fig:temperature_heatmaps}

\end{center}
\end{figure}

Temperature heatmaps revealed a strong seasonal influence, with a maximum of 40°C recorded in January 2024. Both ISO and KNN algorithms identified anomalous peaks in September 2023 and January 2024, though temperature variability remained relatively stable otherwise, suggesting consistent thermal responses.

\paragraph{Comparison of Anomaly Detection Frequencies Across Sensors}
\label{comparison-of-anomaly-detection-frequencies-across-sensors}
\begin{figure}[htbp]
\begin{center}
\includegraphics[width=\linewidth]{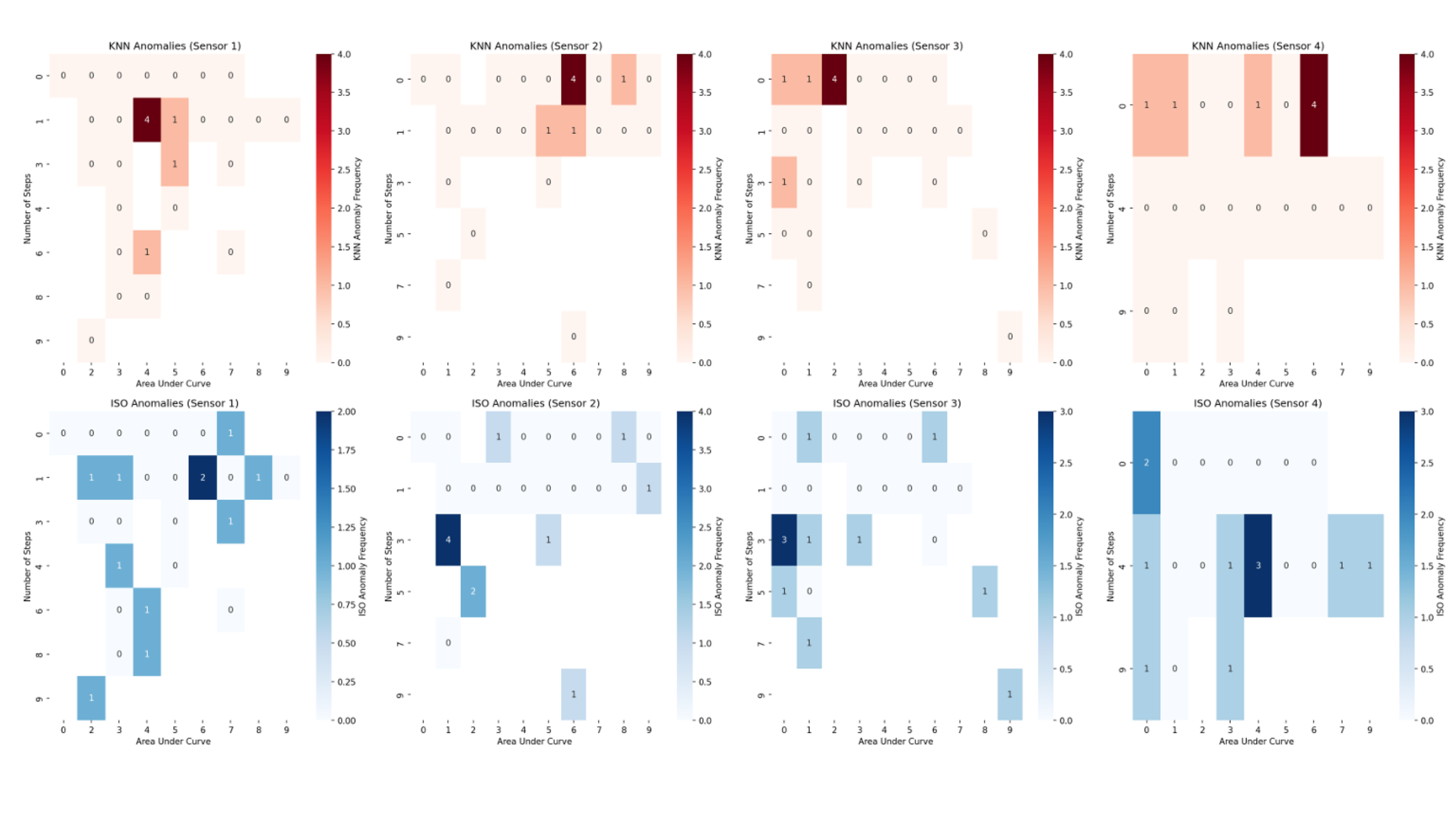}
\caption{Comparison of Anomaly Detection Frequencies Across Sensors}
\label{fig:anomaly_detection_comparison}

\end{center}
\end{figure}

The KNN algorithm tends to identify anomalies in specific, concentrated
areas, whereas the ISO algorithm detects anomalies more broadly across a
wider range of values. This is evident in the results for each sensor,
where KNN shows concentrated anomaly clusters, primarily identifying
anomalies at specific step and AUC values, such as 0 steps and AUC 6 for
Sensor 2, and a narrow range'' for Sensor 3. In contrast, ISO
distributes anomalies more evenly across multiple bins and AUC values.
Overall, KNN is more sensitive to specific pressure and step patterns,
identifying concentrated regions of anomalies with higher frequency,
while ISO has a more distributed approach, indicating a wider definition
of unusual behavior in pressure readings.

\paragraph{Correlation Analysis}
Pearson correlation coefficients were computed between each pressure feature and each temperature feature \textit{across all 312 capture sessions}, yielding aggregated cross-sectional correlations at the session level --- not within-subject or temporal cross-correlations. The highest correlations were observed between mean pressure from sensor~3 (midfoot/heel region) and maximum temperature from sensor~1 ($r = 0.48$), suggesting that sessions with higher sustained midfoot pressure also tended to exhibit elevated foot temperatures. Maximum pressure from sensor~3 and maximum temperature from sensor~1 showed a similar association ($r = 0.41$), providing empirical support for combined multi-modal monitoring consistent with \cite{wilson2024integrating}.

To assess whether combined pressure and temperature 
monitoring provides complementary information beyond what either modality offers 
alone, we examined the overlap between modality-specific anomaly sets detected 
by Isolation Forest across all 312 capture sessions. Of the 17 sessions flagged 
as anomalous by at least one modality, 11 were identified in pressure data only 
(64.7\%), 6 in temperature data only (35.3\%), and notably, zero sessions were 
flagged in both modalities simultaneously. This complete absence of overlap means 
that pressure and temperature anomaly detection are fully complementary under the 
current framework: relying on a single sensor stream would miss 100\% of the 
anomalies detected by the other. This provides strong empirical justification for 
the multi-modal monitoring approach adopted in this study, and suggests that 
pressure and temperature signals capture distinct physiological phenomena rather 
than redundant information. A clinical assessment of whether these modality-exclusive 
anomalies carry distinct pathophysiological significance remains an important 
direction for future work with diabetic patient cohorts.

\paragraph{Seasonal temperature trend}
\label{seasonal-temperature-trend}
\begin{figure}[htbp]
\begin{center}
\includegraphics[width=\linewidth]{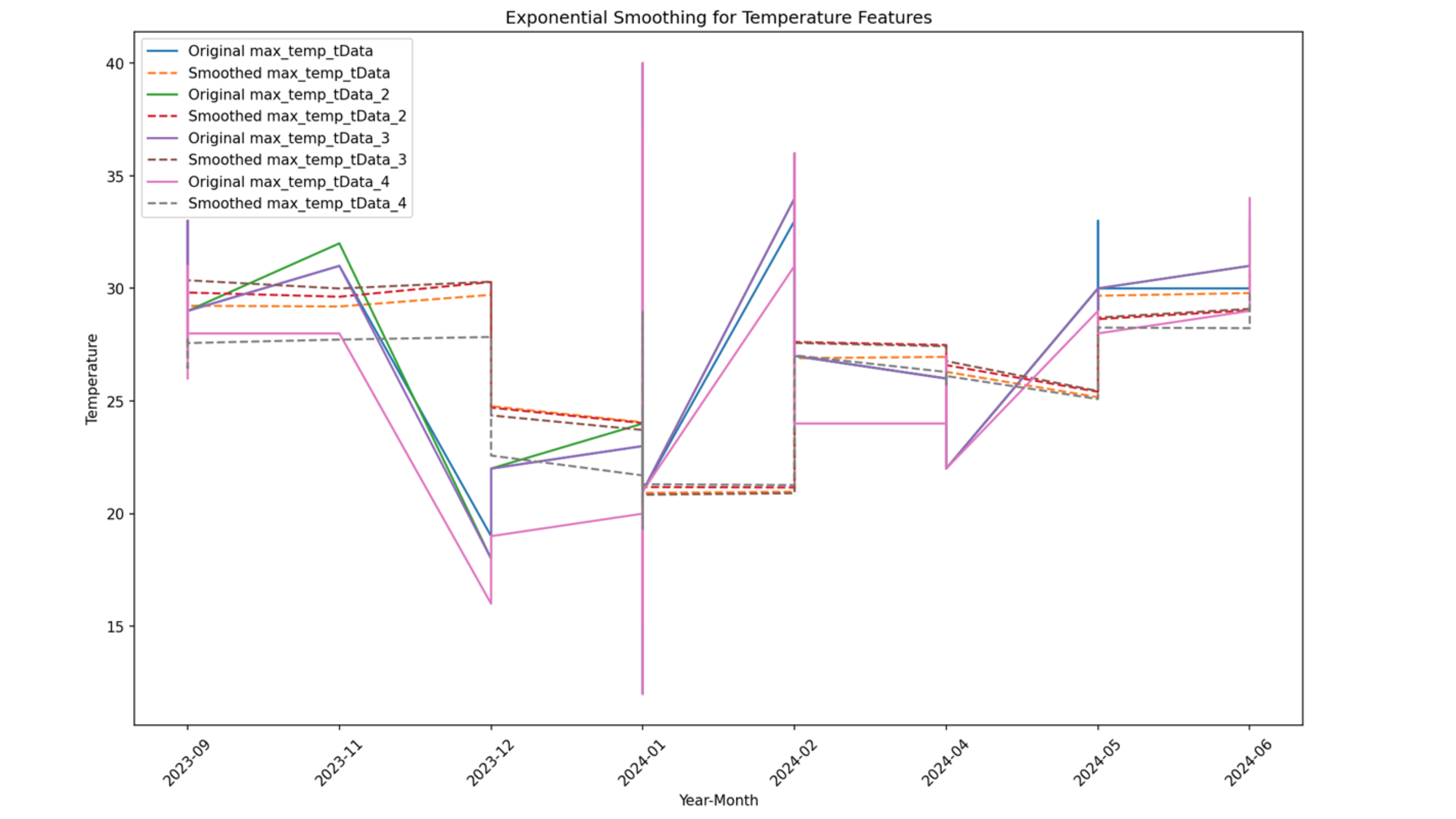}
\caption{Seasonal temperature trend}
\label{fig:seasonal_temperature_trend}

\end{center}
\end{figure}

* The analysis of foot temperature patterns showed a significant drop in
December 2023, potentially indicating reduced circulation or
vasoconstriction due to colder weather, which could increase the risk of
foot ulcers for diabetic patients.

* After the December drop, there was a notable recovery trend in
temperature readings across all sensors, suggesting possible
improvements in blood circulation, overall foot health, or the influence
of increasing ambient temperatures.

* Distinct temperature peaks were observed in January and April 2024,
but these were temporary rather than sustained improvements, emphasizing
the need for continuous monitoring and potential interventions to
maintain improved circulation and reduce ulcer risk.

\section{Results Analysis}
\label{results-analysis}

The findings from this baseline study highlight the 
distinct operational characteristics of the two unsupervised anomaly detection 
algorithms when applied to healthy-subject foot sensor data 
(Figures~\ref{fig:pressureanomalyscores}--\ref{fig:temperatureknnanomalyconcentration}). 
It is important to note that the anomalies detected here represent statistical 
deviations within a healthy cohort they reflect unusual patterns relative to 
the groups normal distribution and should not be interpreted as clinically 
confirmed DFU-related pathology.

Isolation Forest proved more adept at detecting subtle, 
distributed deviations in both temperature and pressure data 
(Figures~\ref{fig:pressureanomalyconcentration},~\ref{fig:temperatureanomalyconcentration}). 
Its partitioning-based mechanism captures a wider range of low-density data 
points, making it sensitive to mild but consistent departures from baseline 
patterns the type of signal that, in a future clinical study with diabetic 
patients, may correspond to early tissue stress or inflammatory changes.

In contrast, KNN/LOF demonstrated higher sensitivity to 
extreme, localised deviations, such as sharp temperature spikes or high-pressure 
events in concentrated foot regions (Figures~\ref{fig:pressureknnanomalyconcentration}, 
\ref{fig:temperatureknnanomalyconcentration}). While effective at flagging the 
most pronounced outliers, this sensitivity resulted in greater inter-method 
disagreement compared to Isolation Forest, flagging additional sessions not 
identified by the latter. In the absence of clinical ground truth, these 
additional detections cannot be confirmed as true or false positives; however, 
the concentration of KNN/LOF-exclusive detections at boundary cases suggests 
lower specificity relative to Isolation Forest under the shared 5 percent contamination 
assumption. This suggests that in a future clinical system, KNN/LOF may be 
better suited as a secondary alarm for severe events rather than as the primary 
continuous monitoring algorithm.

To quantify inter-method agreement, we computed the overlap between anomaly sets identified by Isolation Forest and KNN/LOF across all 312 capture sessions. For pressure data, \textbf{1} session was flagged by both methods (overlap: \textbf{5.9}\%), \textbf{10} by Isolation Forest only, and \textbf{6} by KNN/LOF only; the Jaccard similarity coefficient was \textbf{0.059}, indicating \textbf{low} agreement. For temperature data, \textbf{2} sessions were flagged by both methods (overlap: \textbf{20.0}\%), \textbf{4} by Isolation Forest only, and \textbf{4} by KNN/LOF only; the Jaccard similarity coefficient was \textbf{0.200}, also indicating \textbf{low} agreement. These low agreement values quantify the complementary nature of the two paradigms and provide a reproducible baseline against which future supervised validation can be compared.

The mild positive correlation identified between plantar pressure and foot temperature features (r = 0.41--0.48) provides empirical support for a multi-modal monitoring approach, consistent with existing literature \cite{wilson2024integrating}. This finding suggests that in higher-risk populations, pressure-induced friction and load may be associated with localised thermal changes — a relationship worth exploring directly in diabetic cohorts. Temporally, both algorithms identified anomaly clusters during September--October 2023 and January 2024, periods likely influenced by seasonal environmental conditions such as ambient temperature changes and footwear variations rather than physiological pathology in this healthy sample.

\section{Conclusions}
\label{conclusions}

This study developed and evaluated a baseline anomaly detection framework for continuous foot biomechanics monitoring using time-series temperature and pressure data from wearable sensors, applied to 312 capture sessions of healthy adult subjects. The primary contribution of this work is methodological: it establishes a validated preprocessing pipeline, feature engineering approach, and comparative algorithm evaluation as a necessary foundation for future clinical studies — not a direct DFU diagnostic tool, which would require data from diabetic patients with clinically confirmed outcomes. Two unsupervised machine learning algorithms, Isolation Forest and KNN/LOF, were systematically compared for their ability to detect statistical deviations in multi-modal sensor data. Isolation Forest demonstrated broader sensitivity to subtle, distributed anomalies, while KNN/LOF identified concentrated extreme deviations with greater inter-method disagreement, suggesting complementary roles in a future hybrid monitoring system. A mild positive correlation between pressure and temperature features further supports the rationale for multi-modal sensor integration. Collectively, this work provides a reproducible, open-methodology baseline that future research can build upon when applying these techniques to clinical cohorts of diabetic patients.

\subsection{Limitations}
\label{limitations}

This study presents several important limitations that must be acknowledged. The relatively small dataset collected from healthy adults in a controlled environment fundamentally limits the generalizability of findings to broader diabetic populations \cite{althnian2021impact}. The absence of data from patients with varying stages of diabetic foot disease, different comorbidities, or diverse demographic characteristics constrains the applicability of the developed models to real-world clinical scenarios. Environmental factors such as footwear type, ambient temperature, humidity, and varying activity levels were not incorporated into the analysis, yet these variables significantly influence foot temperature and pressure readings and could lead to false positives or misinterpretation of sensor data in practical applications. The K-Nearest Neighbors/LOF algorithm flagged additional sessions not corroborated by Isolation Forest; in the absence of clinical ground truth, these additional detections cannot be confirmed as true or false positives. However, the concentration of KNN/LOF-exclusive detections at boundary cases suggests lower specificity under the shared 5 percent contamination assumption, which could result in alert fatigue if deployed without refinement in real-time monitoring systems. Additionally, the controlled laboratory setting may not adequately represent the complexities of real-world conditions, including patient compliance variations, sensor placement inconsistencies, and the dynamic nature of daily activities that influence foot biomechanics. Furthermore, clinical translation faces practical hurdles, as adoption of wearable DFU technology can be impeded by concerns about device comfort, reliability, and alert fatigue, which must be addressed for real-world viability.

Furthermore, the absence of clinical ground truth means it is not possible to determine whether the detected anomalies correspond to any DFU-relevant physiological changes; they can only be interpreted as statistical deviations within a healthy population. The dataset's relative homogeneity — healthy subjects recorded under similar controlled conditions — limits the diversity of signal patterns available for model training, and inter-individual variability may be the primary driver of flagged anomalies rather than any latent pathological process.

The database schema did not include individual participant identifiers or demographic information, meaning the exact number of unique participants cannot be determined from the data alone; the 312 capture sessions represent the available unit of analysis. Real-world deployment would additionally need to account for variability in terrain, walking speed, sock thickness, footwear type, ambient humidity, and day-to-day activity levels, none of which were systematically controlled or recorded in this study.

\paragraph{Between-Subject vs.\ Within-Subject Variation}
The absence of participant-level identifiers prevents distinguishing between-subject variation from true anomalies within individuals. Higher plantar pressures or elevated temperatures may simply reflect differences in body weight, foot morphology, or habitual gait patterns across participants rather than pathological deviations. For example, heavier individuals may naturally exhibit higher plantar pressures even with healthy gait mechanics, while variations in foot structure could lead to different baseline temperature patterns. Without within-subject baseline modeling, it is not possible to determine whether a flagged anomaly represents a true deviation from that individual's own normative patterns or simply reflects that individual's position in the population distribution. Future studies should incorporate subject identifiers to enable within-subject baseline modelling, isolating genuine deviations from each participant's own normative patterns rather than from a population-wide distribution.

\subsection{Future Directions}
\label{future-directions}

Future research should prioritize the collection and analysis of larger, more diverse datasets encompassing patients across the spectrum of diabetic foot disease progression to validate and enhance the robustness of the predictive models \cite{althnian2021impact}. The integration of wearable sensors into continuous monitoring systems represents a critical advancement opportunity, where smart insoles or other embedded devices could transmit real-time data to cloud-based platforms for immediate analysis and intervention alerts \cite{anikwe2022mobile, martin2019review}. Model refinement through ensemble learning approaches is a particularly promising direction; this is supported by a growing body of evidence showing that ensemble models are a state-of-the-art technique in DFU image analysis \cite{zhang2022comprehensive} and consistently outperform individual classifiers in predicting DFU risk \cite{zhang2022comprehensive}. Further exploration should include specific advanced architectures, such as recurrent models like Long Short-Term Memory (LSTM) networks, which have demonstrated superior performance in predicting DFU healing progression\cite{spinazzola2025chronic}, and hybrid models that combine deep learning for feature extraction with efficient classifiers like Isolation Forest \cite{spinazzola2025chronic}. The incorporation of contextual environmental variables, including ambient conditions, footwear characteristics, and patient activity patterns, into predictive algorithms could significantly improve model accuracy and clinical relevance \cite{yurur2014context}. Integration of these predictive models into existing clinical workflows through electronic health records and telemedicine platforms would enable real-time anomaly alerts and facilitate timely interventions. Furthermore, the development of context-aware systems that dynamically adjust model parameters based on individual patient characteristics and environmental conditions could enhance the personalization and effectiveness of diabetic foot ulcer prevention strategies \cite{wilden2017iot}. These advancements collectively hold the potential to transform continuous digital monitoring into a standard component of proactive diabetic foot care management.


\bibliographystyle{elsarticle-num}
\bibliography{references}

\end{document}